\documentclass[english,two column]{emulateapj}
\usepackage{bm}
\usepackage{lscape}

\usepackage{graphicx}
\newcommand{\be}{\begin{equation}}
\newcommand{\ee}{\end{equation}}

\newcommand{\ba}{\mbox{\boldmath{$\alpha$}}}
\newcommand{\bt}{\mbox{\boldmath{$\theta$}}}

\begin{document}
\title{Reconstruction of Cluster Masses using Particle Based Lensing
  I: Application to Weak Lensing} 
\author{Sanghamitra Deb, David M. Goldberg \& Vede J. Ramdass} 

\affil{Department of Physics, Drexel
  University, Philadelphia PA-19104} 

\email{sd365@drexel.edu}

\begin{abstract}
We present Particle-Based Lensing (PBL), a new technique for
gravitational lensing mass reconstructions of galaxy clusters.  Traditionally, most
methods have employed either a finite inversion or gridding to turn
observational lensed galaxy ellipticities into an estimate of the
surface mass density of a galaxy cluster.  We approach the problem
from a different perspective, motivated by the success of multi-scale
analysis in smoothed particle hydrodynamics.  In PBL, we treat each 
of the lensed galaxies as a particle and then reconstruct the
potential by smoothing over a local kernel with variable smoothing
scale.  In this way, we can tune a reconstruction to produce constant
signal-to-noise throughout, and maximally exploit regions of high
information density.

PBL is designed to include all lensing observables, including multiple
image positions and fluxes from strong lensing, as well as weak
lensing signals including shear and flexion.  In this paper, however,
we describe a shear-only reconstruction, and apply the method to
several test cases, including simulated lensing clusters, as well as
the well-studied ``Bullet Cluster'' (1E0657-56).  In the former cases,
we show that PBL is better able to identify cusps and substructures
than are grid-based reconstructions, and in the latter case, we show
that PBL is able to identify substructure in the Bullet Cluster
without even exploiting strong lensing measurements. We also make our codes publicly available.
\end{abstract}


\maketitle
\section{Introduction}

Clusters of galaxies are excellent cosmological laboratories
\citep{2007MNRAS.tmp.1149A,2007MNRAS.382..308K,2007MNRAS.tmp.1132M}. For
example, the mass function of clusters is a sensitive probe of
cosmological parameters like $\Omega_m$ and $\sigma_8$
\citep{2007ApJ...657..183R} and its observed evolution is an important
test of theories of structure formation
\citep{1972ApJ...176....1G,2007MNRAS.376..977G,2005MNRAS.360.1393H,2002PhR...372....1C}.
The geometrical shape of cluster Dark Matter halos provide valuable
information on intra-cluster gas distribution
\citep{2005astro.ph..8226F,2007MNRAS.377..883F}.  While simulations
predict central density distribution of matter in clusters to follow
an NFW profile, it is debatable whether observations suggests that
clusters have a central core
(\cite{2003astro.ph..9465S,2006MNRAS.368..518V}). 
  
$\Lambda$CDM structure formation theories also predict that massive
dark matter halos assemble from the hierarchical merging of lower mass
subhalos. As noted by several authors
(\cite{1999ApJ...524L..19M,1999ApJ...522...82K}), the number of
subhalos that survive in N-body simulations is much greater than the
number of dwarf galaxies observed in the Milky Way and the
Andromeda galaxy. On cluster scales, such discrepancies are not observed.
Thus the subhalo mass function in clusters is an important probe of
the CDM theory in this mass scale.

High-resolution, accurate measurements of cluster mass maps
are thus highly desirable.  Gravitational lensing is a powerful tool
to probe the projected mass map of the clusters independent of the
internal dynamics, and has already been widely applied to mapping mass
distribution in clusters \citep{2001ApJ...557L..89W,
  2001ApJ...548L...5H,
  2002ApJ...568..141G,2004MNRAS.353.1176T,
  2005ApJ...621...53B,2007ApJ...666...51L, 2007arXiv0710.2262O,
  2008MNRAS.385.1431H}.  Some researchers
\citep{2004ApJ...617L..13N,2007arXiv0711.4587N} have used the
individual galaxy-galaxy lensing signal to estimate individual galaxy
masses and thus produce a parametric mass reconstruction of the
cluster.  Others have used the weak signal to characterize the overall
potential from the cluster without recourse to parametric models
\citep{1996AAS...189.8206W,1998ApJ...504..636H,2000ApJ...538L.113N,2004IAUS..220..439H}.
%One of the great triumphs of lensing reconstructions of clusters is
%the ``Bullet Cluster'' (1E0657-56), a colliding cluster in which two
%mass peaks were detected
%\citep{2006ApJ...652..937B,2006ApJ...648L.109C} well removed from the
%X-ray gas peaks.

Given the importance of accurately measuring the mass, shape, and
substructure of individual clusters, and given the enormous expense of
long time-exposure observations of clusters, it is extremely important
to maximize the signal-to-noise from a particular dataset, and to produce
high-resolution maps of substructure within individual clusters.
Current mass reconstruction techniques are ill-equipped to handle
multi-scale datasets or clusters with significant clumpiness or
cuspiness, or are jury-rigged to do so.  In this paper, we propose
Particle Based Lensing (PBL; pronounced ``pebble'') as an alternative
approach to cluster reconstruction.

Our outline is as follows.  In \S~\ref{sec:background} we give a brief
review of the essential lensing formalism, and lay out our notation
for the rest of the work.  In\S~\ref{sec:S+W} we describe current
(grid-based) techniques for reconstructing galaxy clusters, and
identify some strengths and complications.  In \S~\ref{sec:particle}
we propose Particle Based Lensing. We then apply this new method to
simple simulated clusters of single and double peak softened isothermal spheres and the ``bullet cluster'' (1E0657-56) in
\S~\ref{sec:applications}.  We conclude in \S~\ref{sec:discuss} with a
discussion of future prospects, including how additional
strong-lensing and flexion information can be incorporated into PBL.
\newpage
\section{Background}
\label{sec:background}
 Before delving into technical details of our method we would
 like to introduce the basic lensing notation to be used throughout
 the paper.  Following \cite{2001PhR...340..291B}, we consider a
 surface mass density $\Sigma({\mathbf\theta})$, where $\mathbf\theta$
 is the angular position in the lens plane. Convergence or
 dimensionless surface mass density is defined as
\begin{equation}
\kappa({\mathbf\theta})={\Sigma({\mathbf\theta}) \over \Sigma_{cr}},
\end{equation}
where
\begin{equation}
\Sigma_{cr}={c^2 \over 4 \pi G}{D_{s} \over D_d D_{ds}}
\end{equation}
$D_{s}$, $D_\mathrm{d}$, and $D_\mathrm{ds}$ are the angular diameter
distances between the observer and the source, the observer and the
lens, and the lens and the source, respectively.  For convenience, we
will define a fiducial critical density for a source plane at
$z_s=z_{ds}=\infty$, and all models will be scaled to this standard.

The convergence is related via a Poisson-like equation to a normalized
potential:
\begin{equation}
\nabla^2 \psi({{\mathbf\theta}})=2\kappa({\mathbf\theta}).
\label{eq:laplace}
\end{equation}
Here and throughout this paper, all derivatives are in angular units
in the lensing plane.  
A single light beam is deflected by:
\begin{equation}
{\mathbf\alpha}=\nabla\psi\ .
\end{equation}
The lens equation relates the source position $\mathbf\beta$ to the
image position(s), ${\bf\theta}$, as:
\begin{equation}
{\beta=\theta-\nabla\psi}.
\label{eq:lenseqn}
\end{equation}
When the lensing potential does not vary appreciably across the
source, the lens mapping can be linearized. The transformation between
the source and the image is given by the Jacobian matrix
\begin{eqnarray}
\nonumber {\bf A}({{\bf \theta}})&\equiv&
\frac{\partial{{\bf \beta}}}{\partial {{\bf \theta}}}=
\left(
\delta_{ij}-\psi_{,ij}\right)\\
&\equiv&
\left(
  \begin{array}{cc}
1-\kappa-\gamma_1 & -\gamma_2 \\
-\gamma_2 & 1-\kappa+\gamma_1
\end{array}
\right)\ .
\label{eq:Adef}
\end{eqnarray}

From this, we see that distortions in shape are well described in
terms of shear which is related to the lensing potentials through the
relations:
\begin{equation}
\gamma_1={1 \over 2}(\psi,_{11}-\psi,_{22})
\label{eq:gam}
\end{equation}
\begin{equation}
\gamma_2=\psi,_{12}
\end{equation}
using Einstein convention for derivatives.

The radial eigenvalue is given by, $\lambda_+=1-\kappa+|\gamma|$ and
the tangential eigenvalue is given by, $\lambda_-=1-\kappa-|\gamma|$ .
The matrix is singular where $\lambda_{\pm}=0$. These points define
the critical curves of the lens.

Third order corrections to the lensing potential becomes
non-negligible when the lensing potential varies across the
image. These observables, the gravitational flexion, were derived in
\cite{2005ApJ...619..741G}. They are more colloquially referred to as
the ``banananess''\citep{2007arXiv0709.1003S} or bending of an
image. It is based on the third angular derivatives of the potential
\citep{2006MNRAS.365..414B} given by,

\begin{eqnarray}
{\cal
  F}&=&(\gamma_{1,1}+\gamma_{2,2})+i\left(\gamma_{2,1}-\gamma_{1,2}\right)\\
\nonumber
&=&\nabla \kappa\\
{\cal G}&=&
  \left(\gamma_{1,1}-\gamma_{2,2}\right)+i\left(\gamma_{2,1}+\gamma_{2,2}\right)
\end{eqnarray}

Each of the deflection ($\ba$), the shear, ($\gamma_1, \gamma_2$), the
convergence ($\kappa$) and the flexion (${\cal F},{\cal G}$), are
linear functions of the potential field.  While the discussion in this
paper primarily focuses on measurements with sources at fixed,
infinite redshift, it should be noted that each of these terms scales
as:
\begin{equation}
\kappa(z_s)=Z(z_d,z_s)\kappa(z_s=\infty)\ .
\end{equation}
and where
\begin{equation}
Z(z_s)=\frac{D_{ds}}{D_s}\ .
\end{equation}
%We discuss the implications of sources at multiple redshifts in
%\S~\ref{sec:global}.

\section{Grid-Based Cluster Lensing}
\label{sec:S+W}
Mass reconstruction studies have been very successful on cluster
scales
(\cite{2001PhR...340..291B,2002A&A...395..385C,2002MNRAS.333..911H,2005ApJ...619L.143B,2007arXiv0710.2262O} and references therein).
Because these systems typically contain many lensed images, the shear
signal can be extracted with high significance.  In this section, we
describe an important class of cluster inversion techniques which
reproduce the convergence field on a grid. 
Our specific choices of grid-based
techniques include those which have already been extended to include
strong-lensing information with non-parametric models and thus provide
a fertile basis for comparison. 
Further, there are many variants even within the sub-category of
grid-based reconstruction techniques.  We focus primarily on their
commonalities, as exemplified by those discussed in
\cite{2005A&A...437...49B,2005A&A...437...39B} and 
\cite{2006A&A...458..349C}. We focus on methods in which
various scalar fields $\{\psi, \kappa\}$ are defined on a Cartesian
grid, and minimized according to the criteria described below.  In so
doing, we note some interesting exceptions: \cite{2005MNRAS.360..477D} and \cite{2001ASPC..237..279S}, who
describe an adaptive mesh technique for refining the field on
different resolution scales and \cite{2002MNRAS.335.1037M,2006MNRAS.372.1289M} who use a variable smoothing scale for their weak lensing mass reconstruction.

\subsection{Weak Lensing on Grids}
\label{subsec:weak}
The standard approach to lensing arclet inversion
\citep{1999A&AS..136..117L} has been to measure the ellipticity of
observed images as an unbiased estimator of the reduced shear:
\begin{equation}
\langle \varepsilon \rangle= g\equiv \frac{\gamma}{1-\kappa}\ .
\label{eq:reducedshear}
\end{equation}
For relatively weak fields ($\kappa \ll 1$), this is very nearly a
direct estimate of the shear, and can perform a direct finite
inversion to estimate the density field.

In recent years, there has been a flurry of work on optimal methods
for non-parametric cluster mass reconstructions
\citep{2005A&A...437...49B,2005A&A...437...39B,2004ApJ...617L..13N}.
In general, these papers focus on estimating the potential
$\{\psi\}$ or convergence $\{\kappa\}$ fields of a cluster by a
$\chi^2$ minimization analysis.  
%2006A&A...458..349C,1998AJ....116.1541A,2005MNRAS.360..477D
Both the shear and the
convergence are linear functions of the potential field.  Thus, if a
model potential field, $\{ \psi \}$, is defined on a grid, then the
shear at some grid-cell, $i$, may be expressed as a linear combination
of potential:
\begin{equation}
\gamma_{1i}=G^{(1)}_{ij}\psi_j
\end{equation}
with a similar expression for the convergence, $\kappa$, and the
imaginary component of the shear, $\gamma_2$.  We refer to these below
simply as $\gamma_i(\{\psi\})$, since we wish to remind the reader
that the estimate of the shear is an explicit function of the test
potential field.  Because these fields are combinations of second
derivatives of the potential field, the $G^{(1)}$ matrix and the
others are easy to compute using finite differencing, and are 
extremely local.  
A very good graphical representation of the finite difference 
operators can be found in \cite{2005A&A...437...39B}.

%\begin{equation}
%\chi^2=\chi^2_W+\chi^2_S+R.
%\label{eq:chi2}
%\end{equation}
%Here $\chi^2_W$ is the contribution from weak lensing as described
%below, $\chi^2_s$ is the strong lensing contribution discussed in \S
%~\ref{subsec:strong} and $R$ is the regularization term explained in
%\S ~\ref{subsec:reg}.

In the weak field limit, the complex ellipticity of a lensed galaxy is
a linear, albeit noisy, estimator of the complex shear field.  The
principle component of noise is the intrinsic ellipticities of
galaxies which follow a Gaussian distribution with standard deviation
of $\sigma_\varepsilon\simeq 0.3$ for each component.
The large variance in intrinsic ellipticities necessitates averaging over many 
images so that the noise in a single grid cell is zero 
or apply an
artificial smoothing scale to a more finely gridded mesh.  For a
weak-lensing only calculation, a $\chi^2$ minimization is performed
on:
\begin{equation}
\chi^2_W=\sum_i \frac{\left( \frac{\gamma_i(\{\psi\})}{1-\kappa_i(\{\psi\}^{(n-1)})}
-\varepsilon_i\right)^2}{\sigma_i^2}
\label{eq:weak_chi2}
\end{equation}
where the estimate of $\kappa$ is taken from the previous iteration of
the potential field, and thus, the model rapidly converges to a
maximum likelihood solution to the potential field.

\subsection{Strong Lensing}
\label{subsec:strong}
A number of researchers, including
\cite{2005A&A...437...49B,2005A&A...437...39B} have noted that a
similar grid-based formalism may be used with strong lensing signals.
Strong+weak (S+W) reconstructions use both shear fields and the
positions of multiply imaged sources can be used to accurately
reconstruct both the cores and halos of clusters.
While our current PBL implementation, described in the next section, does not
currently incorporate strong lensing analysis, we introduce this
component of grid-based lensing reconstructions to illustrate how
directly a strong-lensing analysis could be incorporated into PBL.

Strong lensing by clusters produces an especially elegant result
because if, say, two images are observed at positions, $\bt^A$, and
$\bt^B$, then it must be true that both images originated at the same
(unknown) position in the sky.  Thus, we have a simple relation:
\begin{equation}
\bt^A-\ba(\bt^A)=\bt^B-\ba(\bt^B)
\label{eq:stronglens}
\end{equation}
The appeal of this relationship is that it is fundamentally
linear and thus the angular separation between the two images (itself,
a measurable quantity), can be directly related to the difference in
the first derivatives of the potential at two different points in the
field.

As above, the local derivatives can be computed as:
\begin{displaymath}
\alpha_{xi}=A^{(x)}_{ij}\psi_j
\end{displaymath}
with a similar expression for the y component of the displacement.
The matrix elements of $A$ are easy to compute as they are simply the
1st derivative in a simple grid-based 2nd order difference scheme.
More generally we can express this as $\ba_i(\{\psi\})$.
Thus, an additional $\chi^2$ term can be added:
\begin{equation}
\chi^2_S=\sum_{i,pairs}\frac{\left((\ba^A(\{\psi\})-\ba^B(\{\psi\}))-(\bt^A-\bt^B)\right)^2}{\sigma_i^2}
\end{equation}
and minimized either independently, or simultaneously with the weak
lensing component. 

\subsection{Regularization}
\label{subsec:reg}

Using a $\chi^2$ minimization technique discussed in \S~\ref{subsec:weak} 
it is possible to get a checkerboard pattern due to independent noise
in the two components of ellipticity. This requires the addition of a
regularization term to the $\chi^2$ to suppress this noise.

Scale refinement is also necessary in cases in which
strong+weak lensing signals are combined.  To make this argument
concrete consider a toy isothermal sphere model of a cluster with a
1-d velocity dispersion of 600 km/s.  Each multiply imaged pair will
be separated by twice the Einstein radius, about 20 arc-seconds in
this case. This represents the minimum necessary resolution in the
reconstruction to say anything about strong lensing.

On the other hand, even very efficient space-based weak lensing analysis of
clusters seldom yield more than approximately 100 images/square
arc-minute.  Using a simple Poisson noise estimate, we may achieve
uncertainty of $\sigma_\gamma=0.06$ only with images binned on scales
larger than 30 arc-seconds on a side.  Smaller binnings will naturally
yield larger uncertainties.
Simple grid based method cannot capture both the weak-lensing signal
to high accuracy as well as resolve the strong lensing regime.  In
order to deal with this issue, different investigators have used
different regularization techniques.  

One method is to use a series of finer and finer griddings, and at
each successive level of refinement the convergence field from the
previous level is matched as closely as possible.  The
\cite{2005A&A...437...39B} S+W technique uses this method, with the
weighting parameter selected to provide a $\chi^2$ per degree of
freedom equal to 1, such that:
\begin{equation}
R=\eta \sum_i (\kappa_i^{(n)}-\kappa_i^{(n-1)})^2\ .
\end{equation}
Where $\kappa_i^{(n-1)}$ represents the estimated convergence on the
previous, coarser, gridding, and where $\kappa_i^{(0)}=0$.  We use this
form explicitly in \S\ref{sec:applications} where we test the PBL
method and contrast it to grid-based reconstruction methods.

\subsection{Some Questions}
\label{sec:questions}

Grid-based reconstructions have produced some excellent measurements,
however, there remain a number of complications.  First, grid-based
techniques are really optimized to measure a single scale, the
grid-spacing.  However, as we discuss above, in many interesting
systems, both the structure and information are hierarchical.  An
optimal technique should provide higher resolution in regions of
greater information content.

Moreover, the smoothing and weighting of the strong lensing, weak
lensing, and regularization are created in an {\it ad hoc} basis.  The
ideal smoothing scale should be variable, and such that the
signal/noise ratio of the reconstructed field is similar in every
smoothed cell.

Third, the information from the image ellipticity can only be inverted
outside the critical curves of the lenses.  Inside the (tangential)
critical curve
\citep{1992grle.book.....S,2001stgl.book.....P,1992A&A...260....1S,2004IAUS..220..439H}
there is an abrupt switch in parity of the induced ellipticity of an
image.  More plainly, in the regime $|\gamma| > |1-\kappa|$, the
ellipticity is related to the shear via:
\begin{equation}
\langle \epsilon \rangle_{strong}=\frac{1}{g^\ast}
\label{eq:e_strong}
\end{equation}

As discussed in \S\ref{sec:interp}, this produces a discontinuity in
the ellipticity as a function of $\kappa$ and $\gamma$.  No simple
linear minimization scheme, even an iterative one, will converge to
the ``strong lens'' solution if one starts with a ``weak lens''
initial guess for the local potential field.

\section{Particle Based Lensing  -- PBL}

\label{sec:particle}

In this section, we introduce a new technique called Particle Based Lensing
(PBL) which has the ability to combine the disparate lensing scales in
a coherent way without requiring a regularization scheme.  Several of
the concerns discussed in the previous sections have to do with the
method of discretizing the data for the reconstruction of the lens
potential.  In order to address this, we turn to a technique which is
widely used in another area of astrophysics in which information must
be analyzed on a wide range of physical scales -- numerical N-body
simulations.  Smoothed Particle Hydrodynamics (SPH; see,
e.g. \cite{2005RPPh...68.1703M}, for a recent review) is used in the
modeling of a wide range of physical systems including planets
\citep{2007MNRAS.376.1173W}, star formation
\citep{2003MNRAS.339..312S,2004MNRAS.348..435N} and galaxy formation
\citep{2007MNRAS.375...53K}. The mathematical details of PBL can be complicated, hence we have made our codes for the method public\footnote{\tt http://www.physics.drexel.edu/$\sim$deb/PBL.htm} through our website. 

Before getting into the details, however, it is important to emphasize
what PBL is and is not.  PBL is a new way of discretizing and
describing a reconstructed field.  Moreover, it includes a metric for
comparing a reconstructed model to the observed data.  Everything we
describe below is aimed at demonstrating why this model and metric
are ideal for lensing systems with uneven information content.  While
the current code, and the worked examples are based on weak-lensing
data only, PBL is based on the idea that other probes of the potential
field: strong-lensing positions, flux ratios, and flexion, can be
added to the metric with little complication.

PBL is not, however, a minimization scheme.  That is, much like
grid-based reconstruction methods, PBL fundamentally  consists of a
list of dimensionless potentials and a metric to describe the goodness
of fit.  It does not describe how that minimization criterion is to be
met, however.  In our model section, we describe a number of
approaches to efficient model convergence.  The major argument in
favor of PBL, however, is not that $\chi^2$ minimizes efficiently, but
rather that a low $\chi^2$ in PBL actually corresponds to a model
which closely matches the true underlying system.

\subsection{A Particle Description of the fields}

The fundamental description of the PBL field lies in the a list of
potentials, $\{\psi\}$, one each at the positions of each observed
lensed image.  In order to make the field as continuous as possible,
we may expand the local potential field around the position of any
lensed image, ($\psi_n$, in this case) to arbitrary order:
\begin{equation}
\psi(\bt)=\psi_n+\theta_j\psi_{n,j}+\frac{1}{2}\theta_j\theta_k\psi_{n,jk}+...
\label{eq:potentialtaylor}
\end{equation}
where $\bt$ is the relative offset of the test-point from galaxy
$n$. 

As with grid-based lensing, the local derivatives are composed of a linear combination
of the potentials at each grid point.  That is:
\begin{eqnarray}
\psi_{n,j}&=&D^{(j)}_{nm}\psi_m \\
\psi_{n,jk}&=&D^{(jk)}_{nm}\psi_m
\label{eq:psideriv}
\end{eqnarray}
and so on for arbitrarily higher derivatives.  In reality, we
typically extend the $D^{(\nu)}$ matrices up to 3rd order, where $\nu$
corresponds to 2 matrices for 1st derivatives (displacement field), 3
for second derivatives (shear and convergence), 4 for 3rd derivatives
(flexion).  Here we use Einstein summation convention, the sum over m
runs from 1 to $N_g$.

In terms of the $D^{(\nu)}$ matrices,
equation~(\ref{eq:potentialtaylor}) may be rewritten as:
\begin{equation}
\psi(\bt_m)=\psi_n+\sum_\nu D^{(\nu)}_{nl}X^{(\nu)}_{nm}\psi_l
\end{equation}
where we are explicitly estimating the potential at the m-th galaxy
from the local derivatives defined at the n-th.  We also compactify
equation~(\ref{eq:potentialtaylor}) by defining:
\begin{eqnarray}
X^{(1)}_{nm}&=&\theta_{nx}-\theta_{mx}\\
X^{(2)}_{nm}&=&\theta_{ny}-\theta_{my}\\
X^{(3)}_{nm}&=&\frac{1}{2}\left(\theta_{nx}-\theta_{mx}\right)^2
\end{eqnarray}
and so on.

In order to estimate the derivatives of the potential field near each
galaxy, we need to first compute the $D^{(\nu)}$ matrices.  Since this
problem is under-constrained, we solve for these matrices via a
$\chi^2$ minimization:
\begin{equation}
\chi^2=\sum_m \left( \psi_m-\psi_n-\sum_{\nu,l}
D^{(\nu)}_{nl}X^{(\nu)}_{nm}\psi_l\right)^2 w_{nm}
\label{eq:chi2pbm}
\end{equation}
where $w_{nm}$ is a window function, guaranteeing that only neighboring
galaxies effect the potentials of one
another.  We use a window function of the form:
\begin{equation}
w_{nm}=w\left({|\mathbf\theta_n-\mathbf\theta_m| \over h_n}\right)
\end{equation}
where $h_n$ is inversely proportional to the signal-to-noise at the n-th
image positions. The smoothing scale can also be chosen to be of the 
form $h_{nm}$, i.e symmetric between between the points n and m.

 The signal-to-noise is a function of the local density
of background images and type of constraint (e.g. ellipticity,
positions of multiple images, etc).  A similar approach of using signal
to noise dependent smoothing scale has been used in image analysis of
X-ray data \cite{2006MNRAS.368...65E}.  In regions where there is a
high density of information, the smoothing scale $h_n$ may be set much
lower than in regions of low information density.  

This function must be minimized for every matrix element such that:
\begin{equation}
\frac{\partial \chi^2}{\partial D_{nl}^{(\nu)}}=2\psi_l\sum_m
\left[X^{(\nu)}_{nm}w_{nm}
\left(\psi_m-\psi_n-\sum_{\mu,
  p}D_{np}^{(\mu)}X_{nm}^{(\mu)}\psi_p\right)\right]=0
\label{eq:minD}
\end{equation}
But since equation~(\ref{eq:minD}) is under-constrained, we may also
say:
\begin{equation}
\frac{\partial^2\chi^2}{\partial D_{nm}^{(\nu)}\partial \psi_m}=0
\end{equation}
yielding:
\begin{equation}
\sum_\mu
X_{nm}^{(\nu)}X_{nm}^{(\mu)}w_{nm}D_{nm}^{(\mu)}=X^{(\nu)}_{nm}w_{nm}
\end{equation}
for all $n,m$ and $\nu$.  This can be solved with a simple matrix
inversion, yielding the desired elements for $D^{(\nu)}$.  Of course,
since the elements are a function only of the positions and weightings
of the galaxy images, these elements need only be computed once.  The
method potentially incorporates higher-order derivatives of the
potential, thus, combination of strong, weak and flexion information
becomes a relatively straightforward minimization problem.

\subsection{PBL vs. Regularization}

One of the major advantages of PBL is that we no longer need to
introduce an explicit regularization in order to resolve multi-scale
structure in a reconstruction.  The various regularization schemes
discussed in \S~\ref{subsec:reg} are not motivated from the associated
observations, but are rather derived from assumptions about the mass
profile of a cluster motivated by theory and simulations. 

However, one of the motivations behind using gravitational lensing is
to be able to measure the projected mass without making any
assumptions about the physical state of the system. The advantage of
using PBL is that we do not need to make any assumptions that go into
choosing the regularization term.  The smoothing scale of a ``pebble''
is controlled by $h_n$ which is determined by the local signal to
noise. This means that the position representing weak lensing
measurement will have a low signal to noise and correspondingly a high
$h_n$. This is similar to the typical weak lensing measurement which
is done by averaging over a bin size larger than $\sim
30^{\prime\prime} $. In case of strong lensing we know the positions
of the multiple images for certain, implying high signal to noise and
correspondingly low $h_n$. This can be a few arc-seconds which is the
scale at which the strong lensing structure can be resolved from
multiple images. Thus scales of a few arc-seconds can be combined with
scales greater than $\sim 30^{\prime\prime} $ without making any
assumptions about the mass profile, rather by taking input from the
data.

\subsection{Estimation of the Potential Field}

As with grid-based lensing analysis, in PBL, we use a $\chi^2$
minimization to estimate a maximum-likelihood potential field.  In
this case, however, we sample the potential at every point, and use
the local derivatives of the potentials as defined in
equation~(\ref{eq:psideriv}) to minimize:
\begin{equation}
\chi^2=\sum_{i,n}^{i=2;\ n=N_{gal}}
\left[\frac{\gamma^{(i)}_{n}(\{\psi\})}{1-\kappa_{n}(\{\psi\})}-
\varepsilon^{(i)}_{n}\right]^2 \frac{1}{\sigma_n^2}
\end{equation}
where $i$ ranges from 1,2, and indicates the real or imaginary
component of the shear, reduced shear, or ellipticity.  We shall
henceforth refer to the first term in the parentheses as
$g^{(i)}_n(\{\psi\})$, the estimate of the reduced shear of a model,
and the weighting term outside the parentheses as $w_n$, yielding:
\begin{equation}
\chi^2=\sum_{i,n}
\left[g^{(i)}_n(\{\psi\})-
\varepsilon^{(i)}_{n}\right]^2 w_n
\label{eq:chi2pbl}
\end{equation}
which is the form we will refer to from now on.

This is a weak lensing only expression. Replacing  $g^{(i)}_n(\{\psi\})$ with $1/g^{(i)*}_n(\{\psi\})$ gives the strong lensing counterpart of Eq.~\ref{eq:chi2pbl}. In the next section we discuss how we include this strong lensing version of the equation.

\subsection{Interpolated Ellipticities }

\label{sec:interp}

Linear inversion techniques require that the function to be minimized
is smoothly varying over the domain of interest.  The ellipticities
are given by two functions in the weak and strong lensing regimes by
Eqs.~(\ref{eq:reducedshear},\ref{eq:e_strong}). The boundary of the two
regimes define the critical curves where $|g|=1$ making ellipticities
continuous but not differentiable.

\begin{figure}[h]
\begin{center}
{
\includegraphics[scale=0.4]{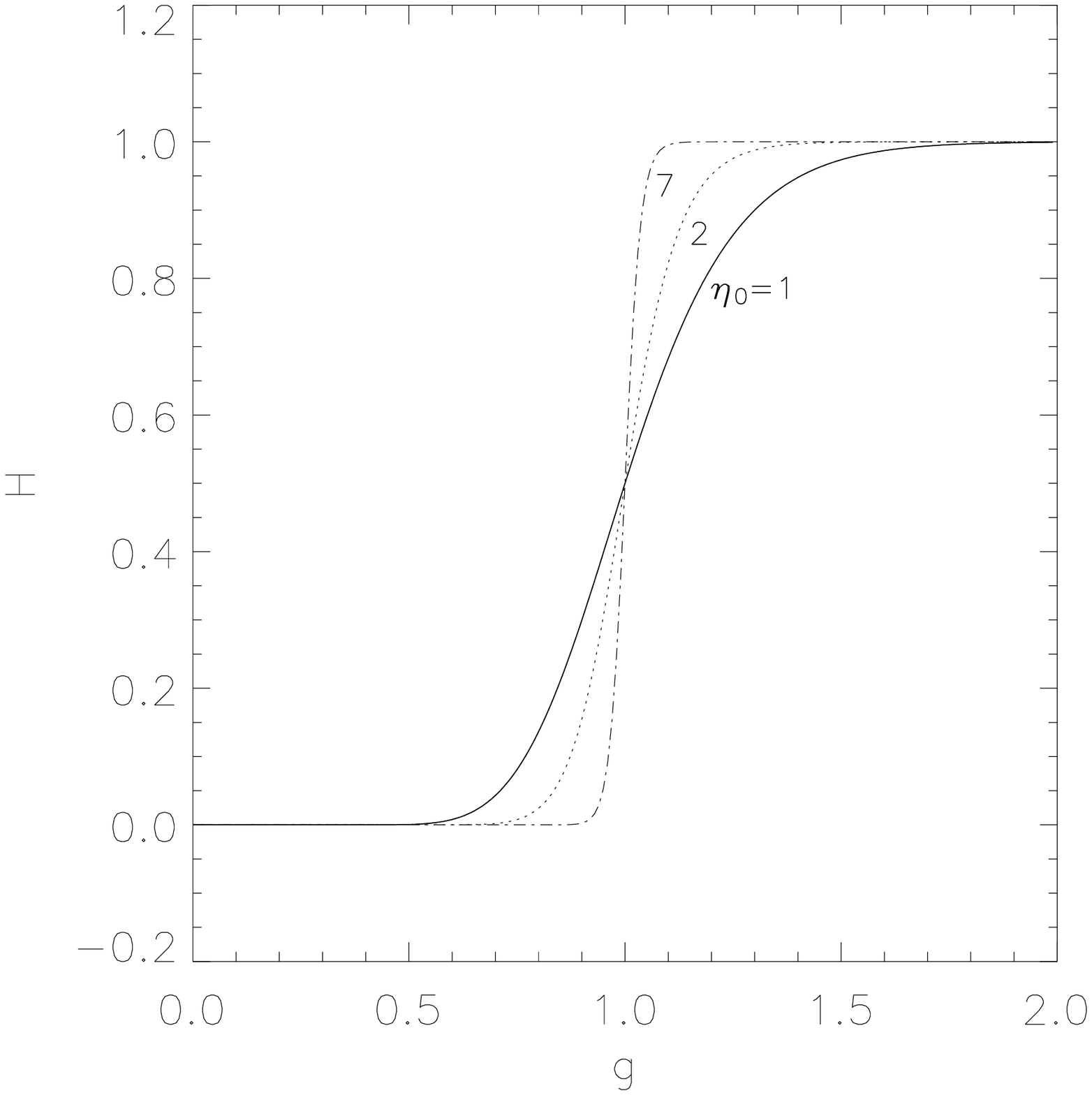}
\includegraphics[scale=0.4]{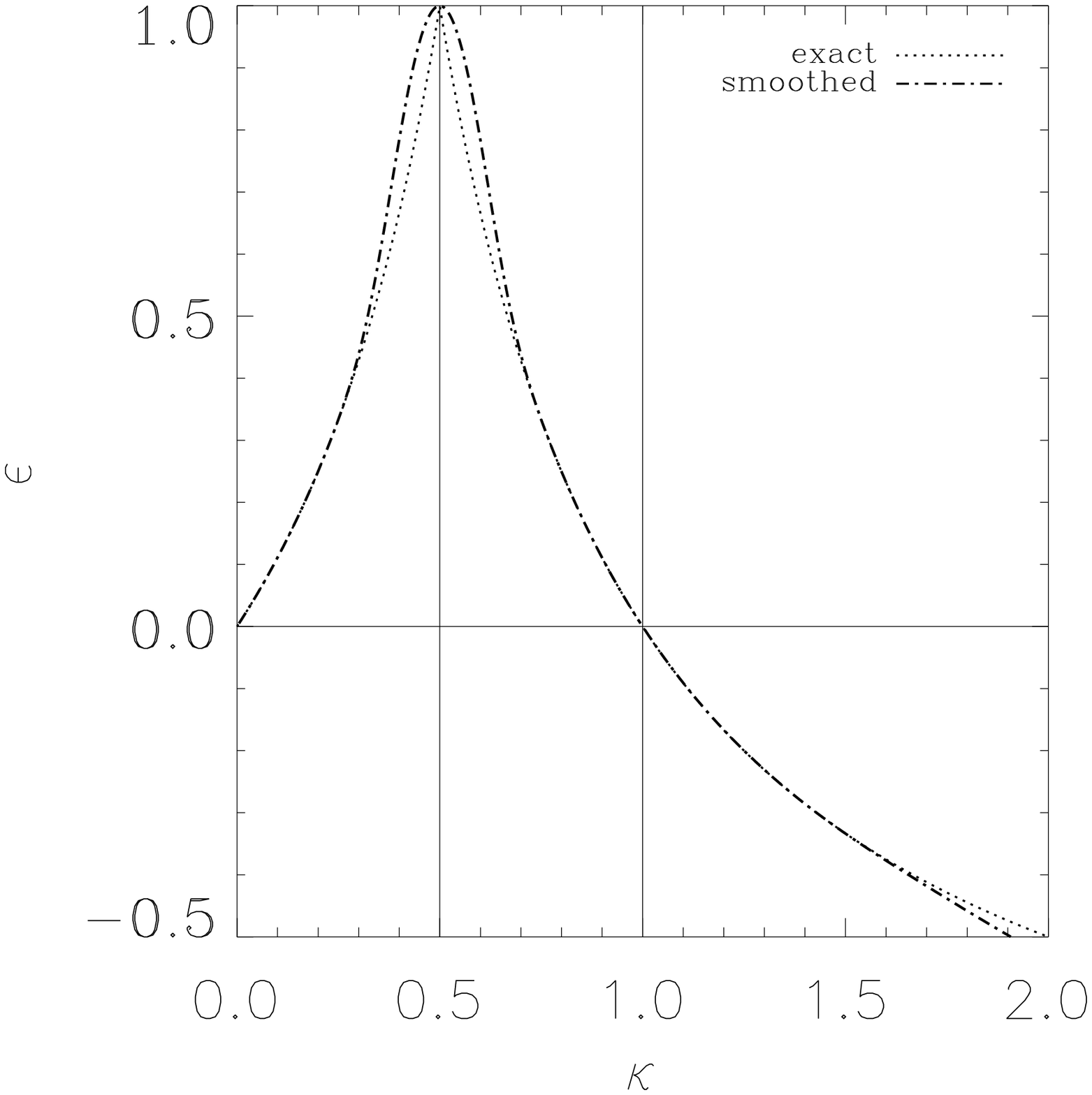}
}
\end{center}
\caption{In the upper panel, we plot the interpolated Heaviside
  step function. It is clear from the plot that the function is only
  approximated by a smooth function near g=1, for all other g it
  behaves like an ordinary step function. Also higher value of the
  parameter $\eta_0$ increases the accuracy. In the lower panel, we
  plot the resulting ellipticity as a function of reduced shear for
  the combination, $|\gamma|=\kappa$. }
\label{fg:etapp} 
\end{figure}

The transition between the two regimes can be facilitated if the
sources are distributed in redshift, but minimization functions will
be much easier if we allow a smoothing of the discontinuities.  This
is a two step process, first we need to write
Eq..~(\ref{eq:reducedshear},\ref{eq:e_strong}) in terms of a step
function,
\begin{equation}
\tilde{\varepsilon}\simeq [1-{\cal H}(g)]g+{{\cal H}(g) \over g^*}
\end{equation}
where the function ${\cal H}(g)$ is a step function at $g=1$.  We may
replace the step function by an approximate smooth function.  We define:
\begin{equation}
u=\eta_0  \left(g^2-{1 \over g^2}\right)
\end{equation}
Here $\eta_0$ is the free parameter that controls the accuracy of the
step function. A higher value of $\eta_0$ makes the step function more
accurate.  The step function is approximated as (Fig.~\ref{fg:etapp}),
\begin{equation}
{\cal H}(u)={1. \over {1+e^{-2 u}}}
\end{equation}

This approximation replaces the ellipticities only in the neighborhood
of the critical curves (discontinuity) by a continuously
differentiable function.  The problem can now be solved by standard
minimization techniques. The interpolated ellipticity function is shown
by a dotted line in the second panel of Fig.~\ref{fg:etapp}, showing
the derivative discontinuity explicitly.

\subsection{$\chi^2$ Minimization}

\label{sec:global}

When we first introduced PBL above, we remarked that it was primarily
a way of describing a lens reconstruction in such a way that a small
$\chi^2$ would necessarily correspond to a good representation of the
underlying field.  In practical terms, though, for a reconstruction
code to be useful, we need to describe a means of minimizing (or
nearly minimizing) the $\chi^2$.  Below, we describe our pipeline for
fast convergence of a maximum likelihood solution.

While PBL is a non-parametric reconstruction scheme, it has the useful
property that we may start a minimization with any assumed model we
like.  However, no extra weight is given to our {\it a priori}
assumptions.  At the end of a minimization we may simply use the
standard techniques to estimate the likelihood of a particular value
of $\chi^2$.  

That said, even with the caveat above regarding smoothing of critical
curves, it is very difficult to smoothly vary a solution such that
strongly lensed regions are produced.  As pointed out by
\cite{2006ApJ...652..937B} a $\chi^2$ minimization process does not
ensure reaching a global minimum.  

To that end, our initial configuration of $\{\psi\}$ is generated by
laying down a small number of Singular Isothermal Spheres (SIS's).
Since there are a low number of parameters (3 for each model sphere),
a global minimum may be reached through a combination of trial and
error, simulated annealing, or even (for small numbers of spheres),
finite sampling.  Indeed, one may even use an interpolation of a
reconstruction recommended by a grid-based solution.  For systems with
strong lenses, one may apply the reconstructed field generated by
``LensPerfect'' \citep{2008arXiv0803.1199C}, for example as a
starting point.

We hasten to remind the reader that while this technique will produce
the optimum parametric fit, it will not, in general, produce the
overall best fit.  As a result, further iteration is required.

We have found that by starting with an initial model with
well-identified strong-lensing regions, convergence to
$\chi^2/DOF\simeq 1$ may be achieved relatively quickly, even if the
strong lensing regions are only approximate.  For the current
implementation of our code, we use Newton's
method to reach a local minimum.  We have found satisfactory, fast,
convergence for several thousand background sources.  

\section{Test Applications}
\label{sec:applications}

In this section, we apply PBL to three systems as a proof of concept.
In the first, we model a Softened Isothermal Sphere, and examine the
relative abilities of PBL and grid-based inversion to reconstruct the
a relatively peaked core.  In the second, we model a superposition of two softened isothermal spheres at a given separation as a simple model of a system with substructure.
Finally, we reconstruct the ``Bullet Cluster'' (1E0657-56) \citep{2002ApJ...567L..27M,2004ApJ...606..819M,2004ApJ...604..596C,2006ApJ...652..937B,2006ApJ...648L.109C}, an
observed multi-peak system of considerable interest.  We show that
using weak lensing alone, we are able to reconstruct both Dark Matter
peaks.

\subsection{Simulation: Softened Isothermal Sphere}

\subsubsection{Model}

We begin by generating a softened isothermal sphere with a potential:
\begin{equation}
\psi=\theta_E\sqrt{\theta^2+\theta_c^2},
\label{eq:softiso}
\end{equation}
and convergence:
\begin{equation}
\kappa=\theta_E{(\theta^2+2 \theta_c^2) \over {(\theta^2+\theta_c^2)}^{3/2}}.
\label{eq:softiso_kap}
\end{equation}
 where $\theta_E$ is the Einstein deflection angle given by $4\pi \left(\frac{\sigma_v}{c}\right)^2 \frac{D_{ds}}{D_s}$.

The data is simulated on a unit square field of view. For simplicity
we have assumed all sources to be at $z=\infty$, with $\theta_E=0.2$
and $\theta_c=0.08$.  We lens 607 background galaxies, and apply an
intrinsic ellipticity (noise) with $\sigma_{e_s}=0.1$ in each of the
principle directions.
For all further calculations we use a $\Lambda$CDM cosmology with $\Omega_m=0.27$ and $\Omega_{\lambda}=0.73$.
This configuration represents a galaxy cluster at a redshift of $z_{lens}=0.4$ with a velocity dispersion of $\sigma_v=850\; \mathrm{km/sec}$. The field of view is $105^{\prime\prime}$ and 0.5 Mpc.

\subsubsection{PBL and Grid-Based Reconstructions}
For the single peak and the double peak simulation(see below), we
perform both a grid-based reconstruction as well as PBL. We use the regularization suggested by
\cite{2005A&A...437...39B}, and described in detail in
\S~\ref{subsec:reg} for the grid based method. 
In case of the single peak the reconstruction is initially performed on a coarse grid ($n_x=6$ gridcells), and is refined up to $n_x=24$, using the $\kappa$ estimated
at each previous step as the prior. For the double peak system we start with $n_x=10$ and refine up to nx=40. For both systems the final reconstruction contains less than one particle per grid cell.

For the PBL reconstruction, we use a smoothing scale of the form:
\begin{equation}
h_n={c \over  (\rho_n)^{\xi}} 
\label{eq:weight}
\end{equation} 
where $\rho_n$ is the local number density of points , $c$ is a
constant, and $\xi$ is a tunable parameter to maximize
signal-to-noise. For our simulation $\xi=1$ is an optimal choice and for the observational case we have used $\xi=0.5$, which is the obvious choice for equalizing signal to noise for every smoothing length. We select $c$ such that the integrated signal-to-noise is greater than unity. The PBL reconstructions are gridded to the same resolution as the grid-based reconstruction to aid visualization. 

For both reconstructions, we begin our iterations with a best-fit
SIS.  We do not, however, use this in the regularization for the
grid-based reconstruction.  

\subsubsection{Results}

Before discussing the results, we note a potential complication.
Gravitational lensing mass measurement suffer from mass sheet
degeneracy when the sources are not distributed in redshift. This
implies that $\kappa$ can be determined up to a degeneracy
$\lambda\kappa+(1-\lambda)$. This transforms to a degeneracy in the
potential of the form,
\begin{equation}
\psi(\bm\theta)\rightarrow\psi^\prime(\bm\theta)=\frac{1}{2}(1-\lambda)\bm\theta^2+\lambda\psi(\bm\theta)
\end{equation}
For the simulated data we have computed the best value of $\lambda$ in
each case and transformed our reconstruction with that value of
$\lambda$ for both the grid based method and PBL.

In Table~\ref{tb:comp} we compare the $\chi^2$ for the best fits of
both the grid-based reconstruction along with PBL for a variety of
smoothing normalization parameters, c. The aim of this table is to quantify the deviation of the reconstructed $\kappa$ from the true $\kappa$. In each case, the ostensible
$\chi^2/\mathrm{DOF}$ is of order unity.  However, one needs to be careful with
simply asserting that the lower $\chi^2$ produces the best result,
since the regularization in grid-based reconstruction adds a penalty
function, and the smoothing scale in PBL lowers the effective degrees
of freedom.

\begin{deluxetable*}{ccccccc}
\tablecolumns{6} \tablewidth{0pc} \tablecaption{Comparison between PBL
  and grid based method.\label{tb:comp}\tablenotemark{a}} \tablehead{ \colhead{Method}
  & \colhead{$\frac{\sum_i^{N_g}(\kappa_i-\kappa_{model,i})^2}{N_g}$}
  & \colhead{$\frac{\sum_i^{N_ {grid}^2}(\kappa_i-\kappa_{model,i})^2
      n_i}{\sum_i^{N_{grid}^2}n_i}$} &
  \colhead{$\frac{\sum_i^{N_{grid}^2}(\kappa_i-\kappa_{model,i})^2
    }{N_{grid}^2}$} & \colhead{$\mathrm{X}^2/\mathrm{DOF}$} & \colhead{$\eta$} &\colhead{c}
}
\startdata
\sidehead{Single Peak}
PBL &0.0200&0.0136&0.0147&1.03&-&0.5\\
PBL &0.0181&0.0128&0.0119&0.94&-&0.7\\
PBL &0.0219&0.0139&0.0131&0.95&-&1.0 \\
PBL &0.0235&0.0140&0.0133&0.94&-&1.3\\
PBL &0.0227&0.0120&0.0121&0.98&-&1.5\\
GRID &0.0311&0.0283&0.0237&0.6&10&-\\
GRID &0.0309&0.0280&0.0223&0.79&30&-\\
GRID &0.0311&0.0280&0.0224&0.94&60&-\\
\cutinhead{Double Peak}
PBL &0.0250&0.0174&0.0167&0.82&-&1.1 \\
PBL &0.0231&0.0168&0.0160&0.80&-&1.38\\
PBL &0.0277&0.0193&0.0180&0.82&-&1.7 \\
PBL &0.0320&0.0219&0.0208&0.87&-&2.0 \\
GRID&0.0570&0.0711&0.0630 &0.92 &20 &-\\ 
GRID&0.0367&0.049&0.039 &0.7 &40 &-\\ 
GRID&0.0359&0.0482&0.0454 &0.83 &60 &-\\ 
\enddata
\tablenotetext{a}{ 
The 2nd, 3rd and the 4th columns represent
  the deviation of the reconstructed $\kappa$ from the true $\kappa$
  weighted  uniformly by galaxy, by local density within gridcells,
  and uniformly by gridcells. Here $\eta$ is the weight given to the 
regularization for grid based method and c is the smoothing
normalization parameter for PBL.}
\end{deluxetable*}
So while both models produce small values of $\chi^2$, the real
question is whether these good fits correspond to a an accurate
reconstruction of the underlying density field.  In
Table~\ref{tb:comp}, we do several comparisons which relate the
reconstructed $\kappa$ at each galaxy (or grid-point) with the 
true $\kappa$ modeled by the simulation for both the single peak 
and the double peak. The comparisons are 
done with a range of values for both $\eta$ (the regularization 
weight in grid based method) and $c$ (the proportionality constant 
in PBL). 

The first column in Table 1 describes the method used, i.e 
either PBL or the grid based method. The second column describes the 
difference between reconstructed $\kappa$ and the true $\kappa$ for every 
galaxy position.In order to extract this information from the gridded 
reconstruction we have used the nearest grid point method which simply 
means the $\kappa$ at each galaxy position is assigned the value at the 
corresponding grid cell. The third column describes the deviation of the 
reconstructed $\kappa$ from the true $\kappa$ at every grid cell weighted 
by the number of image galaxies in that grid cell. The 4th column describes 
the difference between reconstructed and true $\kappa$ weighted uniformly 
over the grids.  In each of the 3 comparisons,
PBL reproduces the original reconstruction with the highest fidelity. 
The 5th column gives the $\chi^2/\mathrm{DOF}$, the 6th gives the
regularization parameter $\eta$ for grid based method and the 7th
column gives the smoothing normalization parameter for PBL.

\begin{figure*}
\begin{center}
\includegraphics[scale=0.7]{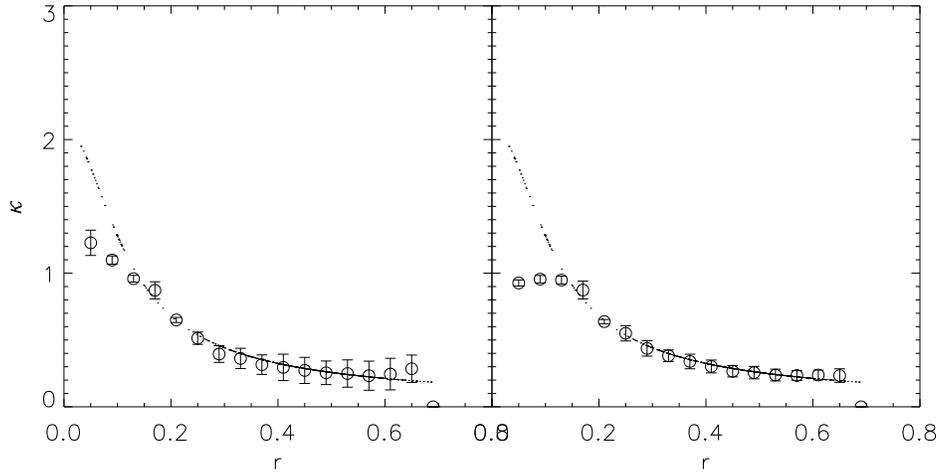}
\caption{A radial plot of the reconstructed convergence ($\kappa$) of a
  simulated Softened Isothermal Sphere.  The circles represent binned
  reconstructed $\kappa$ and the error bars represent the scatter in
  each bin. The dots represent the true value of $\kappa$ given by
  Eq.~\ref{eq:softiso_kap}. The x-axis represents the radial distance from the center of the 
 softened isothermal sphere. The radial distance is scaled and hence unit-less.
Upper Panel: Using PBL. Lower panel: Using grid
  based method. The error bars in the radial plot using PBL is
  higher. This is because the errors introduced in PBL are dependent on
  the local signal to noise which are not spherically symmetric. In
  the grid based method the errors are averaged uniformly on the
  length scale of a single grid which makes the radial scatter very
  low.}
\label{fg:kap_recon_rad}
\end{center}
\end{figure*}

In Fig.~\ref{fg:kap_recon_rad}, we show the radial reconstruction of
the softened isothermal sphere using the two different techniques.  
The bulk of the
penalty associated with the grid-based reconstruction relative to PBL
occurs near the core. By construction,
PBL is designed to perform well in this regime. 

\subsection{Simulation: A Double Peaked Cluster}

\subsubsection{Model}

While PBL has been shown to perform well modeling a single Softened
Isothermal Sphere in the previous section, the other major goal of
this method is to reconstruct small-scale substructure in a system.
To that end, we model a doubly-peaked system with 814 lensed
background galaxies.  As before, they are placed on a unity grid, and
are modeled as 2 Softened Isothermal Spheres, with:
$x_1=0.65,y_1=0.35$, $x_2=0.35,y_2=0.65$, $\theta_{E1}=0.2,\theta_{c1}=0.1$, 
$\theta_{E2}=0.2$, and $\theta_{c2}=0.1$.  The simulated noise, and
reconstruction technique for the double peaked system are identical to
the single peak system. 
This is system of two sub-clusters at a redshift of $z_{lens}=0.4$ having a velocity dispersion of $\sigma_v=850\;km/sec$ separated by 226 kpc. The field of view is $105^{\prime\prime}$.

\subsubsection{Results}

As with a single sphere, both PBL and grid-based reconstructions
produce $\chi^2/\mathrm{DOF}\simeq 1$, as illustrated in Table~\ref{tb:comp}.
However, as with the single sphere reconstruction, PBL produces
smaller errors with regards to the underlying model than does the
grid-based reconstruction.  

\begin{figure}[h!]
\includegraphics[scale=0.4]{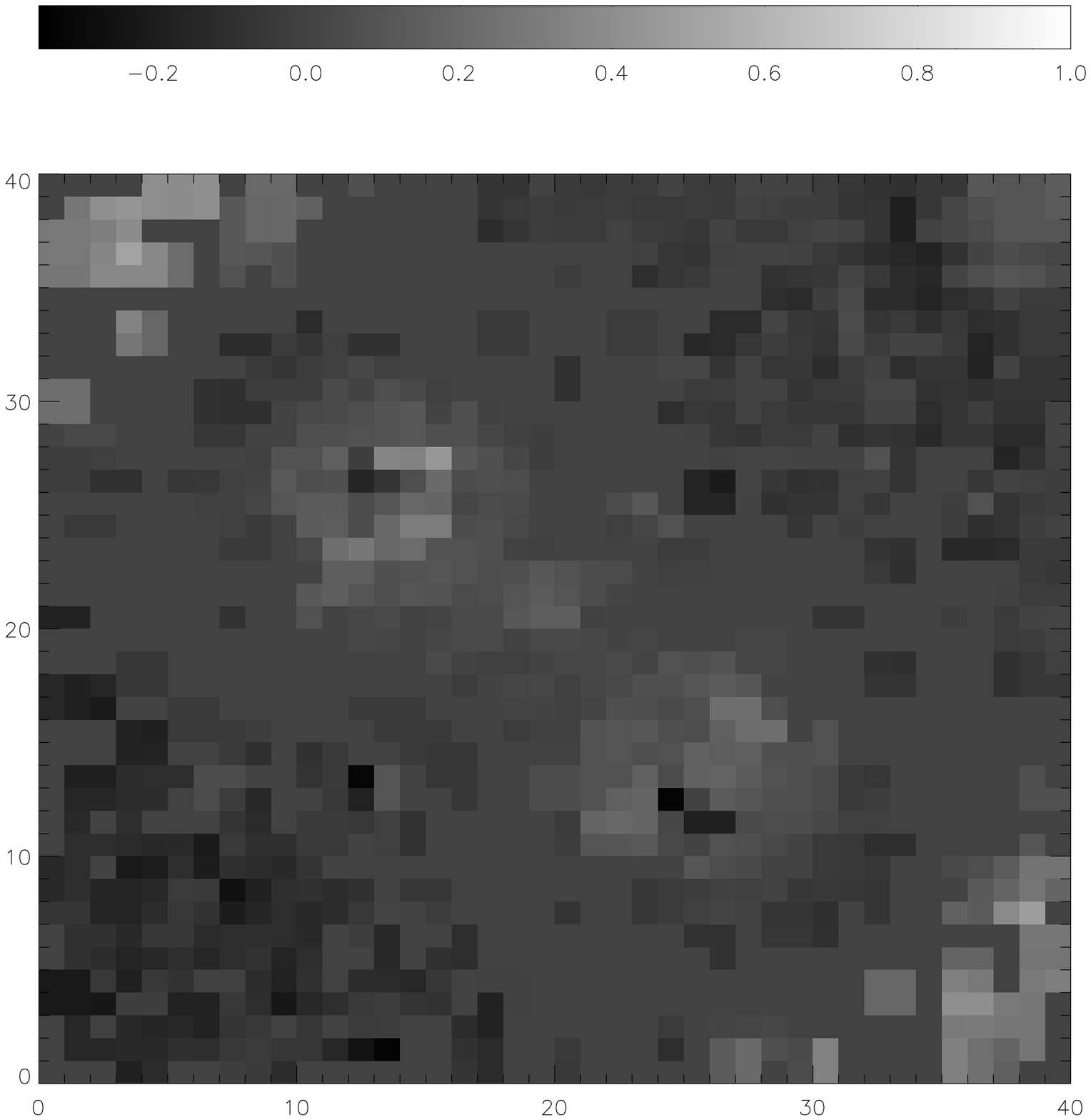}
\includegraphics[scale=0.4]{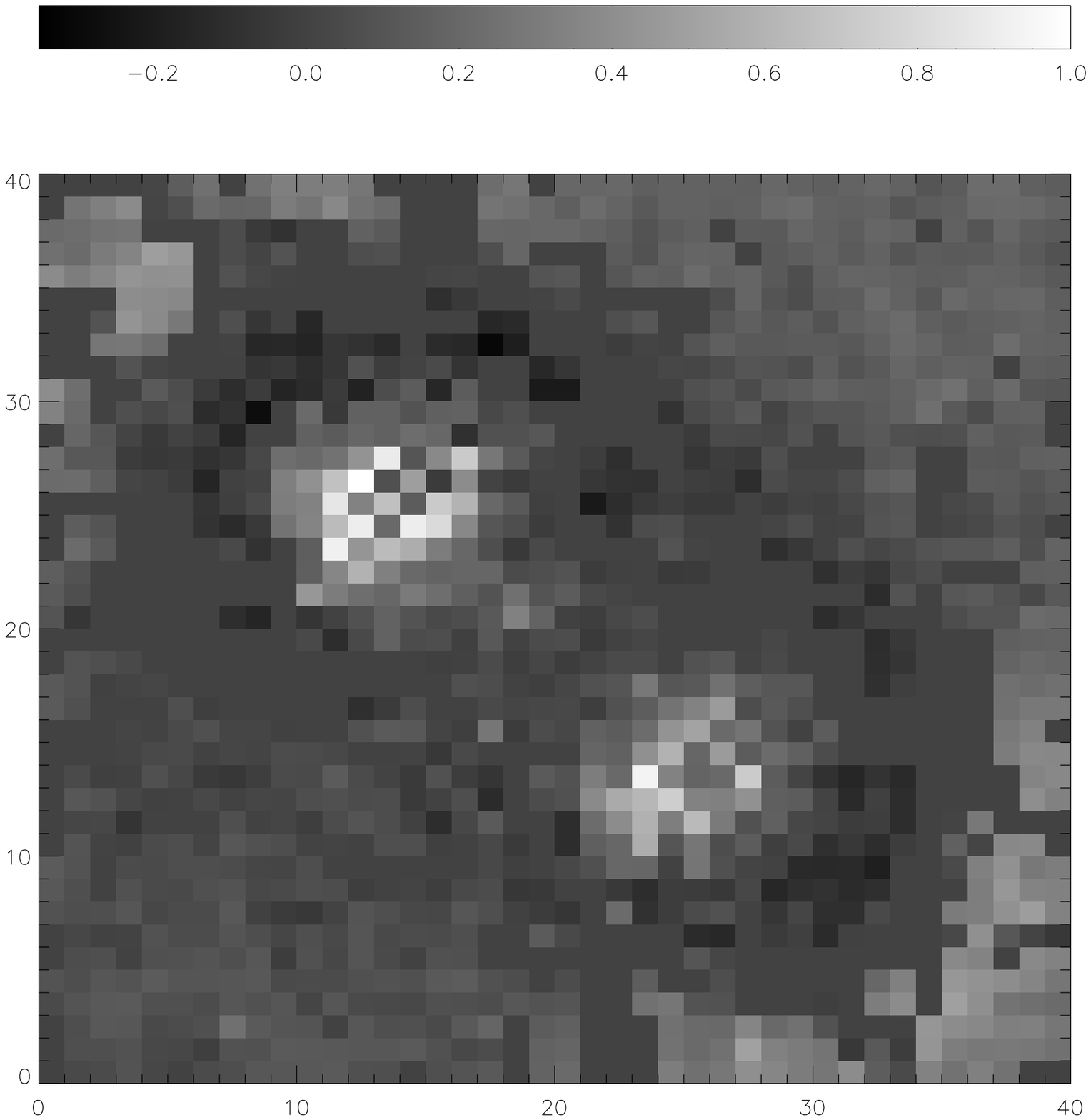}
\caption{ The plot of the difference between reconstructed convergence,$\kappa$
  and true $\kappa$ for the double peak SIS system. Left Panel: Using PBL
  . Right panel: Using grid based method described in
  \S~\ref{sec:S+W}. Both maps are gridded for easy visualization. 
Also there are empty grid cells with no image galaxies. The value for 
those grid cells in the above difference map is set to zero for both 
reconstructions. As we can see the error in the cores of the 
peaks is much lower using PBL mass reconstruction}
\label{fg:kap_recon_image_2peak}
\end{figure}

In Fig.~\ref{fg:kap_recon_image_2peak}, we show a grey-scale plot of
the residuals between the underlying model and each of the
reconstructions.  Unsurprisingly, both models have the greatest
difficulty reproducing highly peaked cores, though PBL is more
responsive to high local gradients in $\kappa$. We describe the general quality of the fit in Table~\ref{tb:comp}.

\subsection{Observation: The Bullet Cluster}

\subsubsection{Observations}

Finally we perform a mass reconstruction of the bullet cluster
(1E0657-56). This galaxy cluster is a rare supersonic merger in the
plane of the sky. Its distinctive structure and orientation makes it
an ideal cluster for observing dark matter using gravitational
lensing. It consists of two sub-clusters separated by 0.72 Mpc,  
which have just undergone a
merger and are moving away from each other. The western sub-cluster is less massive and the eastern main cluster is more massive. The line-of-sight velocity difference suggest that their cores passed each other ~100 Myr ago. The collisionless dark
matter in each of the sub-clusters have crossed each other but the
fluid-like intracluster plasma is in the process of electromagnetic and thermal
interaction producing high X-ray luminosity far removed from lensing
mass peaks \citep{2006ApJ...648L.109C,2006ApJ...652..937B}.

For the Bullet Cluster, we perform a PBL reconstruction only, since it
has been well-studied with grid-based methods (using \cite{1995A&A...302..639S,1995ApJ...439L...1K}) and the $\kappa$-contours 
are publicly available.  We use publicly
available weak lensing data from the Bullet Cluster Project
Page\footnote{\tt
  http://flamingos.astro.ufl.edu/1e0657/public.html}. The catalog was
constructed using data from three different instruments: the ESO/MPG
Wide Field Imager, IMACS on Magellan, and two pointings of ACS on HST.
The shapes of the galaxies were measured independently on each of the
image sets averaging for the common galaxies. The weighting for each galaxy is based on its significance of detection in every image set and normalized appropriately \citep{2006ApJ...648L.109C}. 

The catalogs were combined using weighted average reduced shear measurements and the weights of individual galaxies were increased when they occurred in several catalogs. This weighting is listed in the shear catalog. We include this weighting in our reconstructions as well and choose only those images 
with a weighting greater than 1. 
As we have already illustrated in the simulations PBL is most effective when 
the information density is variable, i.e close to the core of the clusters. 
In case of the bullet cluster we zoom into a region bounded by $104.53^o$ to $104.69^o$ 
in right ascension and $55.92^o$ to $55.97^o$ in declination. Following this cut, our sample includes
1259 weak lensing background galaxies. In order to do the mass reconstruction we use the average redshift of this sample, $z=0.91$.

\subsubsection{Reconstruction}

The Bullet Cluster was made famous by the direct detection of dark matter by \cite{2006ApJ...648L.109C}.   Indeed, since one
of the major findings of this group is that the dark matter appears
offset from X-ray emissions, we do not include any prior model when
reconstructing the system, but are able to achieve fast convergence with two clearly visible peaks.
This reconstruction guides us in choosing an initial condition for subsequent 
$\chi^2$ minimization. 

We have calculated the integrated mass within 150 kpc of
  each peak. The main peak has a mass of $1.57\times10^{14}M_\odot$
  and the sub-cluster has a mass of $0.9\times10^{14}M_\odot$.
  \cite{2004ApJ...604..596C} report a value of $(1.02 \pm 0.16)\times
  10^{14}M_\odot$ for the main peak and $(0.66\pm0.19)\times
  10^{14}M_\odot$ for the sub-cluster within 150 kpc of the each peak.
  In each case, our estimate exceeds that of Clowe et al. by
  approximately 3.4 $\sigma$.  However, more a more recent S+W
  reconstruction by the same group \citep{2006ApJ...652..937B} yields
  masses of $(2.8\pm0.2)\times 10^{14}M_\odot$ around the main peak
  and $(2.3\pm0.2)\times 10^{14}M_\odot$ around the sub-cluster within
  250 kpc of each peak.  Inclusion of strong lensing information makes
  reconstruction of the cores more accurate and also leads to a higher
  estimates of the mass.  Even correcting for the greater area, this
  suggests Clowe's initial mass estimate may have been low.

Our mass estimates using PBL is higher than the weak lensing
reconstruction of \cite{2004ApJ...604..596C}, and thus more in line
with the S+W results.  This is a result of a difference in method.
For example, we start from an initial condition and iterate to the
correct solution whereas \cite{2004ApJ...604..596C} have fitted a
radially averaged shear profile to the NFW or King profile. As already
seen in the simulations using an initial condition recovers values of
$\kappa$ close to the core with greater accuracy. This implies that
while most weak lensing $\kappa$ maps report $\kappa$-contours less
than 1 using initial condition and PBL we are able to get $\kappa$
greater than 1. This implies that the mass we measure will also be
greater than the typical weak lensing mass measurement. Also to
measure the mass of the sub-cluster \cite{2004ApJ...604..596C} have
removed the mass of the main cluster to avoid over-estimation of the
mass, we have not considered this effect in our reconstruction.

\begin{figure}[h!]
\includegraphics[scale=0.38]{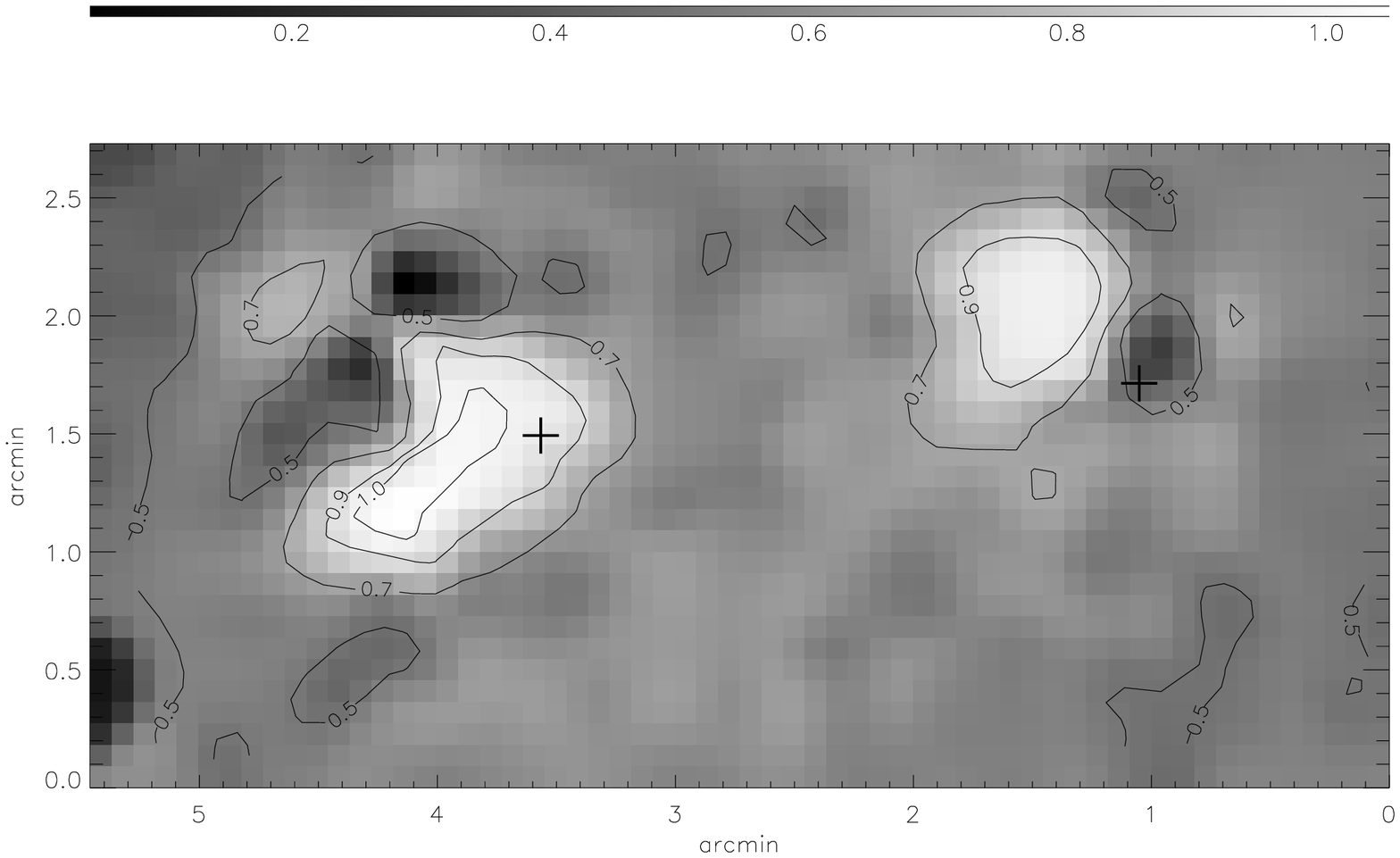}
\includegraphics[scale=0.38]{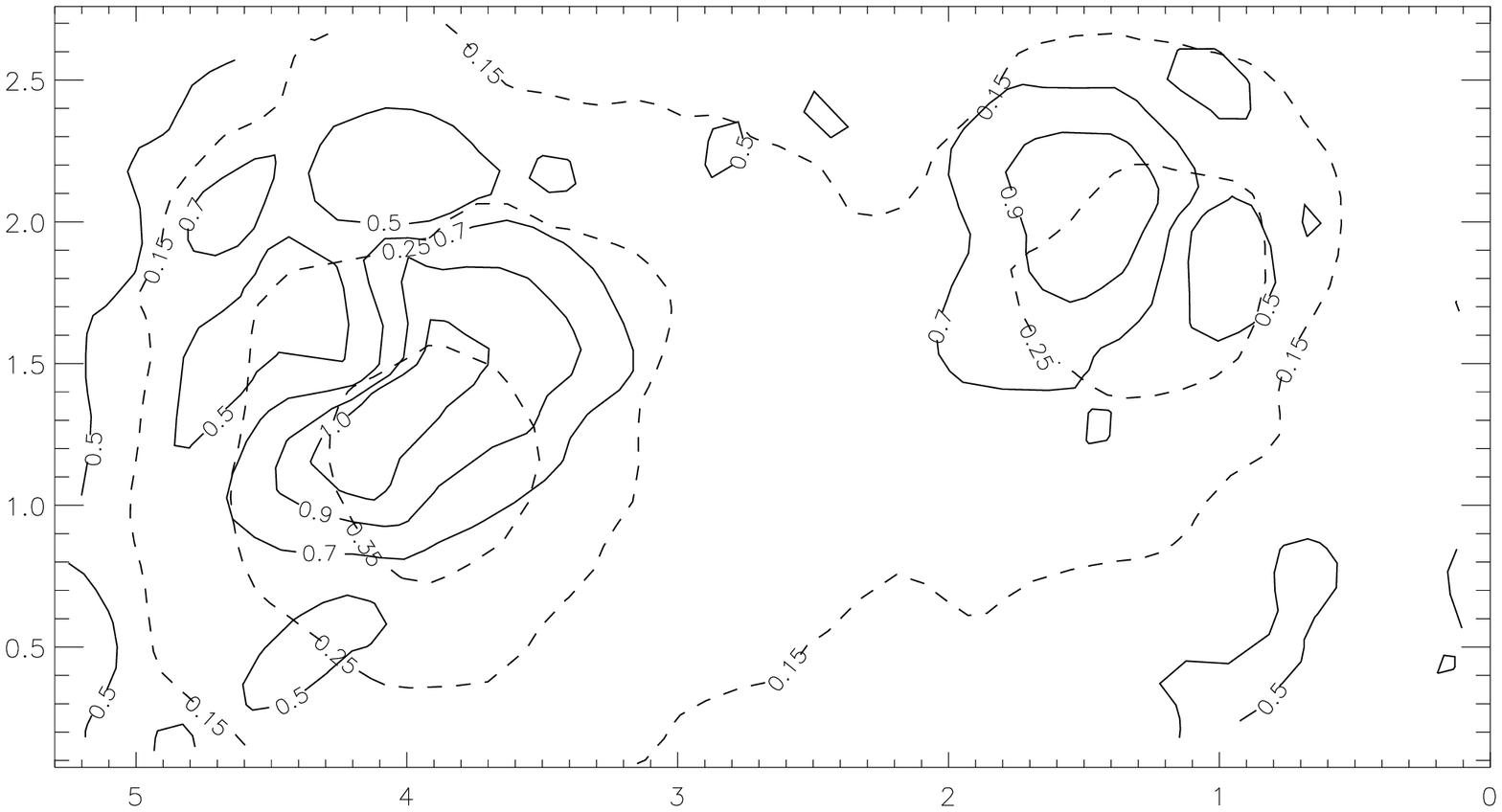}
\caption{A weak-lensing only reconstruction of the bullet cluster
  using PBL described in \S~\ref{sec:particle}.  Note that both
  substructure peaks are clearly identified. Upper Panel: This the
  $\kappa$-map using PBL. The cross denotes the centroid of the
  multiply imaged positions. Lower Panel: This a comparison of the
  $\kappa$ contour derived using PBL(solid) and the publicly available
  contour plot of $\kappa$(dashed). }
\label{fg:bullet_new}
\end{figure}

In Fig.~\ref{fg:bullet_new}, we show our PBL reconstruction of the
bullet cluster.  Note that, despite using weak lensing signals only,
we are able to identify both density peaks and using initial
conditions we are able to get $\kappa>1$ for the main peak. We also 
do a comparison of the publicly available $\kappa$-contour with the 
$\kappa$-contours reconstructed using PBL. The location of the main 
peak coincides for both reconstruction. The sub-cluster contours for 
PBL are slightly removed from the publicly available $\kappa$-contours.

Error analysis for PBL will be discussed in detail in future papers. In particular the noise covariance matrix,$\langle(\kappa-\langle\kappa\rangle)(\kappa-\langle\kappa\rangle)^T\rangle$, will give us important insights into the errors caused by the reconstruction method. A bootstrap method can also be used to determine error bars on mass measurements from observations. In case of simulations several Monte Carlo realizations of the noise can be used to study the errors.

\section{Discussion and Future Prospects}
\label{sec:discuss}

\subsection{Additional Signals}
Thus far, we have developed the formalism for PBL, and done worked
examples demonstrating how it may be applied to weak-lensing
reconstruction.  It is designed to model structure hierarchically, in
part because of the great success of Strong+Weak lensing analysis.

Several groups have already shown how multiple image positions may be
added to the information yielded by lens ellipticities to produce
very high quality mass-maps of clusters.  It was our desire to
maximally exploit the different information scales of the strong and
weak lensing signals which motivated the development of PBL in the
first place.  

However, there is yet more information besides image differences
potentially available which may be utilized in a reconstruction.
Consider that in addition to the two constraints generated by the
positional difference between two images, we also can measure a flux
ratio, and 2 ellipticity differences.  Thus, in principle, we have 5
measurable, model parameters per strong lensing pair rather than 2,
and in an idealized case, this improves potential resolution of a
system in the strong lensing regime by $\sqrt(5/2)\simeq 1.6$.

As a way of guiding the future development of PBL, we discuss possible
future avenues of investigation below.

\subsubsection{Flux}
Apart from the centroid position, the Petrosian flux of an image is
the most straightforward to measure.  The relationship between
magnification lens is simply the inverse of the determinant of the
projection matrix:
\begin{equation}
\mu=\frac{1}{(1-\kappa)^2-|\gamma|^2}
\end{equation}
Unlike the displacement vectors ({\ba}), which are simple linear
operators of the potential field (the gradient), or the weak-lensing
shear field which is nearly so (since in the limit of $\kappa << 1$,
the image ellipticity is an unbiased estimator of the shear field),
the flux is a highly nonlinear function of the shear and convergence
fields.  This accounts, in part, for the reason that it has not been
used previously in cluster reconstructions. Here we note that
\cite{2007ApJ...663...29S} show that the image positions itself
constrain the fluxes for a source with three non-collinear components.
This is a special case, for cluster lensing three component sources
for strong lensing may not always be available. Also,
\cite{2007MNRAS.376..180N} use magnification information in their
parametric mass modelling of clusters.

The other major consideration is that magnification is not a smoothly
varying function of the potential fields.  It is well-known that on the
critical curves, magnification goes to infinity (see,
e.g. \cite{1992grle.book.....S} for an extensive discussion), but this
is a set of measure zero, so in and of itself produces no problem.
The issue is that the parity of the image reverses as an image that
crosses the critical curve.

Negative magnification means nothing more than reversal of image
parity, and thus cannot generally be easily detected.  Thus, we are
much more interested in computing terms which scale like $\mu^2$.
Indeed, since we cannot measure the magnification directly, but only
the flux, we propose that the combination:
\begin{equation}
\frac{\mu_A^2-\mu_B^2}{\mu_A^2+\mu_B^2}=\frac{f_A^2-f_B^2}{f_A^2+f_B^2}
\end{equation}
is directly measurable, and has no poles.  

\begin{figure}[h]
\includegraphics[scale=0.4]{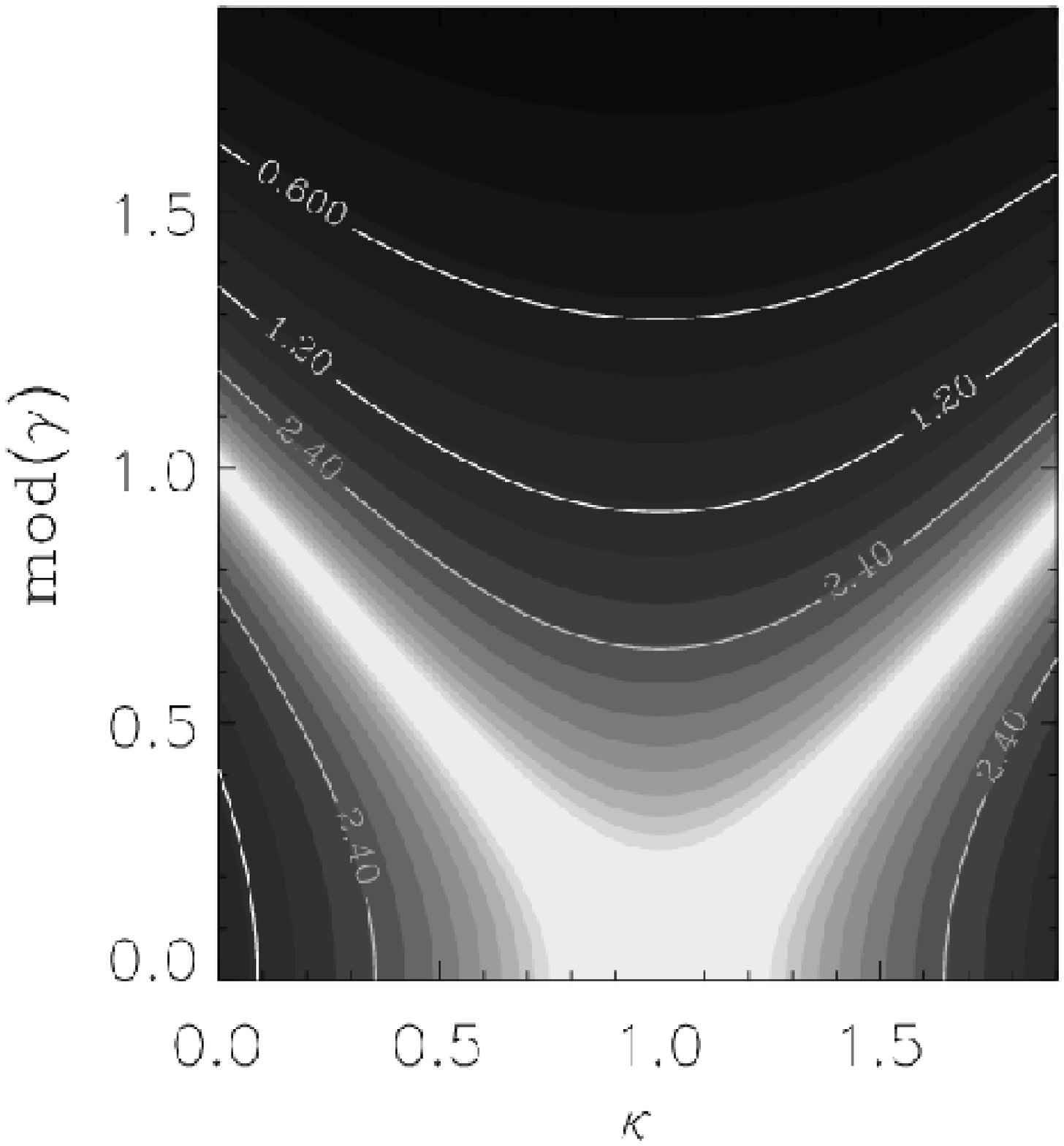}
\includegraphics[scale=0.4]{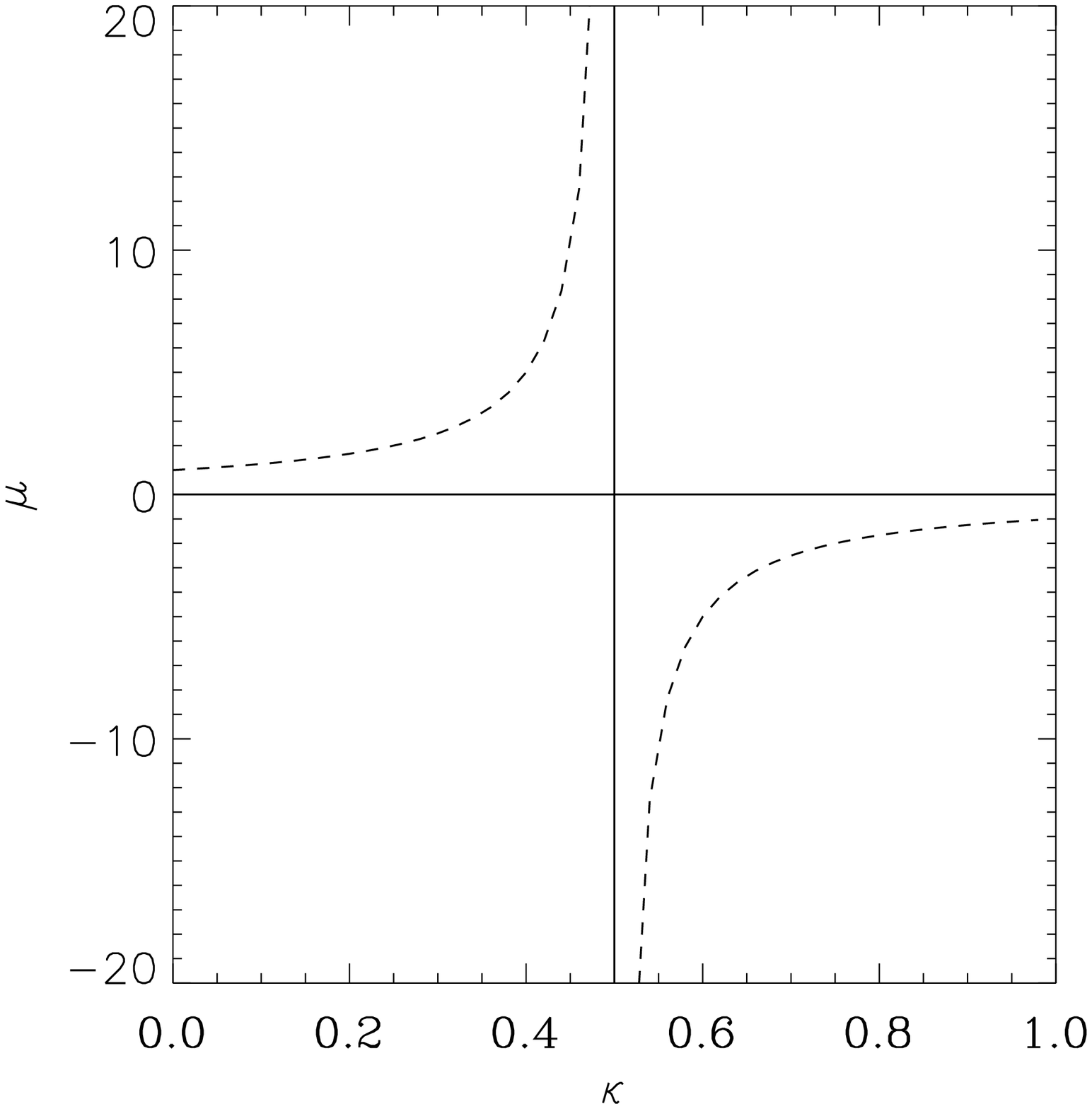}
\caption{The magnification as a function of shear and convergence.
  The lower panel is a simple slice through the upper, with the choice
  $\gamma=\kappa$.  Neither the magnification nor its derivatives are a
  continuous function.  Moreover, flux ratios are only measurable for
  systems with at least two images (obviously).  One or more of the
  images will necessarily have negative parity.  Thus, a solution
  to the potential field which is found using standard relaxation
  methods will not normally converge to a negative parity estimate for
  any magnification.}
\label{fg:mu_parity}
\end{figure}

Even so, a lensing model {\it predicts} a parity for a particular
image, and as with ellipticity, minimization, there is a discontinuity
in the derivatives.  In Fig.~\ref{fg:mu_parity} we show the
magnification (including sign) as a function of convergence and
shear. 

\subsubsection{Ellipticity Differences}

Likewise, while most measurements of the shear are based on an
assumption that any {\it given} image is randomly oriented, two images
of the same source are not.  The difference in their measured
ellipticity can be wholly modeled by the relative lensing fields at
their respective locations.  If both images were in the weak regime,
we would be able to use the simple estimator
\begin{equation}
\varepsilon_A-\varepsilon_B=\simeq \gamma_A-\gamma_B
\end{equation}
where all terms in the equation are complex, and thus provide two 
constraints with high signal to noise per image pair.  

In general, however, a more likely configuration is that one image may
be in the strong regime, and one in the weak.  If we can determine
from the configuration of lenses which is which, we might imagine a
better estimator as:
\begin{equation}
\varepsilon_A-\varepsilon_B=\frac{1}{g^\ast_A}-g_B
\end{equation}
with the only associated noise corresponding to photon noise rather
than random variance in the intrinsic ellipticity of the images.  

\subsubsection{Flexion}

Thus far, the analysis of clusters in the weak or semi-weak regime has
primarily relied on shear.  However, recently,
\cite{2007arXiv0710.2262O}, and \cite{2007ApJ...666...51L} have worked
on reconstructing A1689 using flexion.  In particular, the Okura group
used a Fourier inversion suggested by \cite{2007arXiv0709.1003S}.
However, the advantage of our proposed PBL is that flexion (and, in
principle, {\it any} higher-order derivative of the potential) may be
explicitly included as additional constraints in the cluster
reconstruction.  Unlike Fourier techniques, which rely on binning of
the data, the PBL method will allow us to exploit the natural
small-scale signal probed by flexion.

\subsection{Summary}

We have developed PBL, a new particle based technique of mass
reconstruction of clusters. The distinguishing feature of PBL is its ability to adjust its smoothing scale depending on the local signal to noise or the type of constraint and thus not require any regularization. PBL has the scope of calculating derivatives up to any order. Hence, lensing constraints that are a function of the derivatives of the potential can be easily included in the reconstruction. 
In this paper we have successfully applied PBL to do weak lensing only mass reconstruction for a single peak and a double peak system. We have made the codes for PBL publicly available for application weak lensing measurements through our website(see \S~\ref{sec:particle}). The codes have been tested on the data sets and simulations described in the paper. A larger data sample will require modification of the current version of codes.

As already explained PBL is a method of discretizing data and not a minimization method. A $\chi^2$ minimization does not necessarily ensure reaching a global minimum. In many cases the global minimum is guarded by steep walls surrounded by shallow valleys.
Without any prior knowledge of the mass distribution it is very easy to get trapped in a shallow valley and not reach the global minimum. We have started with an initial condition and interpolated the ellipticity function to aid us in this regard. 
 
In future work we will be including the additional constraints, like the flux
ratios, ellipticity differences and flexion along with measured
ellipticities and strong lensing positions. We will also be exploring different minimization schemes to facilitate convergence to a global minimum.

\section{Acknowledgments}
The authors would like to thank for useful discussions with Mike
Jarvis, Bhuvnesh Jain and Kevin Olson. We would also like to thank Marusa Brada{\v
  c} and Douglas Clowe for providing us with the ``Bullet Cluster''
(1E0657-56) data. DMG gratefully acknowledges discussions with Richard
Massey.  SD would like to thank Ben Metcalf, Prasenjit Saha and Sudeep
Das for valuable comments. We would also like to thank the referee for thoughtful suggestions. This work was supported by NASA ATP
NNG05GF61G.

\newpage
 
% LocalWords:  Sanghamitra Vede Ramdass Drexel ellipticities ellipticity NFW et
% LocalWords:  triaxial baryonic CDM subhalos subhalo virial intra arXiv lensed
% LocalWords:  hoc flexion al Okura banananess cr Jacobian Clowe Hoekstra SEF
% LocalWords:  Broadhurst arclet Luppino arclets Bradac Natarajan Springel PBL
% LocalWords:  Cacciato Abdelsalam discretized supercritical Petters SPH jk nm
% LocalWords:  discretizing Woolfson Hernquist Nagamine Kaufmann Peretto nd
% LocalWords:  cuspy Petrosian BGRT

%\bibliography{clusters}

\begin{thebibliography}{68}
\expandafter\ifx\csname natexlab\endcsname\relax\def\natexlab#1{#1}\fi

\bibitem[{{Abdelsalam} {et~al.}(1998){Abdelsalam}, {Saha}, \&
  {Williams}}]{1998AJ....116.1541A}
{Abdelsalam}, H.~M., {Saha}, P., \& {Williams}, L.~L.~R. 1998, \aj, 116, 1541

\bibitem[{{Allen} {et~al.}(2007){Allen}, {Rapetti}, {Schmidt}, {Ebeling},
  {Morris}, \& {Fabian}}]{2007MNRAS.tmp.1149A}
{Allen}, S.~W., {Rapetti}, D.~A., {Schmidt}, R.~W., {Ebeling}, H., {Morris},
  R.~G., \& {Fabian}, A.~C. 2007, \mnras, 1149

\bibitem[{{Bacon} {et~al.}(2006){Bacon}, {Goldberg}, {Rowe}, \&
  {Taylor}}]{2006MNRAS.365..414B}
{Bacon}, D.~J., {Goldberg}, D.~M., {Rowe}, B.~T.~P., \& {Taylor}, A.~N. 2006,
  \mnras, 365, 414

\bibitem[{{Bartelmann} \& {Schneider}(2001)}]{2001PhR...340..291B}
{Bartelmann}, M., \& {Schneider}, P. 2001, \physrep, 340, 291

\bibitem[{{Brada{\v c}} {et~al.}(2006){Brada{\v c}}, Clowe, {Gonzalez},
  {Marshall}, {Forman}, {Jones}, {Markevitch}, {Randall}, {Schrabback}, \&
  {Zaritsky}}]{2006ApJ...652..937B}
{Brada{\v c}}, M., Clowe, D., {Gonzalez}, A.~H., {Marshall}, P., {Forman}, W.,
  {Jones}, C., {Markevitch}, M., {Randall}, S., {Schrabback}, T., \&
  {Zaritsky}, D. 2006, \apj, 652, 937

\bibitem[{{Brada{\v c}} {et~al.}(2005{\natexlab{a}}){Brada{\v c}}, {Erben},
  {Schneider}, {Hildebrandt}, {Lombardi}, {Schirmer}, {Miralles}, {Clowe}, \&
  {Schindler}}]{2005A&A...437...49B}
{Brada{\v c}}, M., {Erben}, T., {Schneider}, P., {Hildebrandt}, H., {Lombardi},
  M., {Schirmer}, M., {Miralles}, J.-M., {Clowe}, D., \& {Schindler}, S.
  2005{\natexlab{a}}, \aap, 437, 49

\bibitem[{{Brada{\v c}} {et~al.}(2005{\natexlab{b}}){Brada{\v c}}, {Schneider},
  {Lombardi}, \& {Erben}}]{2005A&A...437...39B}
{Brada{\v c}}, M., {Schneider}, P., {Lombardi}, M., \& {Erben}, T.
  2005{\natexlab{b}}, \aap, 437, 39

\bibitem[Broadhurst et al.(2005)]{2005ApJ...621...53B} Broadhurst, T., et 
al.\ 2005, \apj, 621, 53
 
\bibitem[{{Broadhurst} {et~al.}(2005{\natexlab{b}}){Broadhurst}, {Takada},
  {Umetsu}, {Kong}, {Arimoto}, {Chiba}, \& {Futamase}}]{2005ApJ...619L.143B}
{Broadhurst}, T., {Takada}, M., {Umetsu}, K., {Kong}, X., {Arimoto}, N.,
  {Chiba}, M., \& {Futamase}, T. 2005{\natexlab{b}}, \apjl, 619, L143

\bibitem[{{Cacciato} {et~al.}(2006){Cacciato}, {Bartelmann}, {Meneghetti}, \&
  {Moscardini}}]{2006A&A...458..349C}
{Cacciato}, M., {Bartelmann}, M., {Meneghetti}, M., \& {Moscardini}, L. 2006,
  \aap, 458, 349

\bibitem[{{Clowe} {et~al.}(2006){Clowe}, {Brada{\v c}}, {Gonzalez},
  {Markevitch}, {Randall}, {Jones}, \& {Zaritsky}}]{2006ApJ...648L.109C}
{Clowe}, D., {Brada{\v c}}, M., {Gonzalez}, A.~H., {Markevitch}, M., {Randall},
  S.~W., {Jones}, C., \& {Zaritsky}, D. 2006, \apjl, 648, L109

\bibitem[{{Clowe} {et~al.}(2004){Clowe}, {Gonzalez}, \&
  {Markevitch}}]{2004ApJ...604..596C}
{Clowe}, D., {Gonzalez}, A., \& {Markevitch}, M. 2004, \apj, 604, 596

\bibitem[{{Clowe} \& {Schneider}(2002)}]{2002A&A...395..385C}
{Clowe}, D., \& {Schneider}, P. 2002, \aap, 395, 385

\bibitem[{{Coe} {et~al.}(2008){Coe}, {Fuselier}, {Benitez}, {Broadhurst},
  {Frye}, \& {Ford}}]{2008arXiv0803.1199C}
{Coe}, D., {Fuselier}, E., {Benitez}, N., {Broadhurst}, T., {Frye}, B., \&
  {Ford}, H. 2008, ArXiv e-prints, 803

\bibitem[{{Cooray} \& {Sheth}(2002)}]{2002PhR...372....1C}
{Cooray}, A., \& {Sheth}, R. 2002, \physrep, 372, 1

\bibitem[{{Diego} {et~al.}(2005){Diego}, {Protopapas}, {Sandvik}, \&
  {Tegmark}}]{2005MNRAS.360..477D}
{Diego}, J.~M., {Protopapas}, P., {Sandvik}, H.~B., \& {Tegmark}, M. 2005,
  \mnras, 360, 477

\bibitem[{{Ebeling} {et~al.}(2006){Ebeling}, {White}, \&
  {Rangarajan}}]{2006MNRAS.368...65E}
{Ebeling}, H., {White}, D.~A., \& {Rangarajan}, F.~V.~N. 2006, \mnras, 368, 65

\bibitem[{{Falco} {et~al.}(1985){Falco}, {Gorenstein}, \&
  {Shapiro}}]{1985ApJ...289L...1F}
{Falco}, E.~E., {Gorenstein}, M.~V., \& {Shapiro}, I.~I. 1985, \apjl, 289, L1

\bibitem[{{Flores} {et~al.}(2005){Flores}, {Allgood}, {Kravtsov}, {Primack},
  {Buote}, \& {Bullock}}]{2005astro.ph..8226F}
{Flores}, R.~A., {Allgood}, B., {Kravtsov}, A.~V., {Primack}, J.~R., {Buote},
  D.~A., \& {Bullock}, J.~S. 2005, ArXiv Astrophysics e-prints

\bibitem[{{Flores} {et~al.}(2007){Flores}, {Allgood}, {Kravtsov}, {Primack},
  {Buote}, \& {Bullock}}]{2007MNRAS.377..883F}
---. 2007, \mnras, 377, 883

\bibitem[{{Giocoli} {et~al.}(2007){Giocoli}, {Moreno}, {Sheth}, \&
  {Tormen}}]{2007MNRAS.376..977G}
{Giocoli}, C., {Moreno}, J., {Sheth}, R.~K., \& {Tormen}, G. 2007, \mnras, 376,
  977

\bibitem[{{Goldberg} \& {Bacon}(2005)}]{2005ApJ...619..741G}
{Goldberg}, D.~M., \& {Bacon}, D.~J. 2005, \apj, 619, 741

\bibitem[{{Gray} {et~al.}(2002){Gray}, {Taylor}, {Meisenheimer}, {Dye}, {Wolf},
  \& {Thommes}}]{2002ApJ...568..141G}
{Gray}, M.~E., {Taylor}, A.~N., {Meisenheimer}, K., {Dye}, S., {Wolf}, C., \&
  {Thommes}, E. 2002, \apj, 568, 141

\bibitem[{{Gunn} \& {Gott}(1972)}]{1972ApJ...176....1G}
{Gunn}, J.~E., \& {Gott}, J.~R.~I. 1972, \apj, 176, 1
\bibitem[Heymans et al.(2008)]{2008MNRAS.385.1431H} Heymans, C., et al.\ 
2008, \mnras, 385, 1431 

\bibitem[{{Hoekstra} {et~al.}(2001){Hoekstra}, {Franx}, {Kuijken}, {Carlberg},
  {Yee}, {Lin}, {Morris}, {Hall}, {Patton}, {Sawicki}, \&
  {Wirth}}]{2001ApJ...548L...5H}
{Hoekstra}, H., {Franx}, M., {Kuijken}, K., {Carlberg}, R.~G., {Yee}, H.~K.~C.,
  {Lin}, H., {Morris}, S.~L., {Hall}, P.~B., {Patton}, D.~R., {Sawicki}, M., \&
  {Wirth}, G.~D. 2001, \apjl, 548, L5

\bibitem[{{Hoekstra} {et~al.}(1998){Hoekstra}, {Franx}, {Kuijken}, \&
  {Squires}}]{1998ApJ...504..636H}
{Hoekstra}, H., {Franx}, M., {Kuijken}, K., \& {Squires}, G. 1998, \apj, 504,
  636

\bibitem[{{Hoekstra} {et~al.}(2002){Hoekstra}, {Franx}, {Kuijken}, \& {van
  Dokkum}}]{2002MNRAS.333..911H}
{Hoekstra}, H., {Franx}, M., {Kuijken}, K., \& {van Dokkum}, P.~G. 2002,
  \mnras, 333, 911

\bibitem[{{Hoekstra} {et~al.}(2004){Hoekstra}, {Yee}, \&
  {Gladders}}]{2004IAUS..220..439H}
{Hoekstra}, H., {Yee}, H.~K.~C., \& {Gladders}, M.~D. 2004, in IAU Symposium,
  Vol. 220, Dark Matter in Galaxies, ed. S.~{Ryder}, D.~{Pisano}, M.~{Walker},
  \& K.~{Freeman}, 439--+

\bibitem[{{Horellou} \& {Berge}(2005)}]{2005MNRAS.360.1393H}
{Horellou}, C., \& {Berge}, J. 2005, \mnras, 360, 1393

\bibitem[{{Kaiser}(1995)}]{1995ApJ...439L...1K}
{Kaiser}, N. 1995, \apjl, 439, L1

\bibitem[{{Kaufmann} {et~al.}(2007){Kaufmann}, {Mayer}, {Wadsley}, {Stadel}, \&
  {Moore}}]{2007MNRAS.375...53K}
{Kaufmann}, T., {Mayer}, L., {Wadsley}, J., {Stadel}, J., \& {Moore}, B. 2007,
  \mnras, 375, 53

\bibitem[{{King}(2007)}]{2007MNRAS.382..308K}
{King}, L.~J. 2007, \mnras, 382, 308

\bibitem[{{Klypin} {et~al.}(1999){Klypin}, {Kravtsov}, {Valenzuela}, \&
  {Prada}}]{1999ApJ...522...82K}
{Klypin}, A., {Kravtsov}, A.~V., {Valenzuela}, O., \& {Prada}, F. 1999, \apj,
  522, 82

\bibitem[{{Leonard} {et~al.}(2007){Leonard}, {Goldberg}, {Haaga}, \&
  {Massey}}]{2007ApJ...666...51L}
{Leonard}, A., {Goldberg}, D.~M., {Haaga}, J.~L., \& {Massey}, R. 2007, \apj,
  666, 51

\bibitem[{{Luppino} {et~al.}(1999){Luppino}, {Gioia}, {Hammer}, {Le F{\`e}vre},
  \& {Annis}}]{1999A&AS..136..117L}
{Luppino}, G.~A., {Gioia}, I.~M., {Hammer}, F., {Le F{\`e}vre}, O., \& {Annis},
  J.~A. 1999, \aaps, 136, 117

\bibitem[{{Mannucci} {et~al.}(2007){Mannucci}, {Maoz}, {Sharon}, {Botticella},
  {Della Valle}, {Gal-Yam}, \& {Panagia}}]{2007MNRAS.tmp.1132M}
{Mannucci}, F., {Maoz}, D., {Sharon}, K., {Botticella}, M.~T., {Della Valle},
  M., {Gal-Yam}, A., \& {Panagia}, N. 2007, \mnras, 1132

\bibitem[{{Markevitch} {et~al.}(2004){Markevitch}, {Gonzalez}, {Clowe},
  {Vikhlinin}, {Forman}, {Jones}, {Murray}, \& {Tucker}}]{2004ApJ...606..819M}
{Markevitch}, M., {Gonzalez}, A.~H., {Clowe}, D., {Vikhlinin}, A., {Forman},
  W., {Jones}, C., {Murray}, S., \& {Tucker}, W. 2004, \apj, 606, 819

\bibitem[{{Markevitch} {et~al.}(2002){Markevitch}, {Gonzalez}, {David},
  {Vikhlinin}, {Murray}, {Forman}, {Jones}, \& {Tucker}}]{2002ApJ...567L..27M}
{Markevitch}, M., {Gonzalez}, A.~H., {David}, L., {Vikhlinin}, A., {Murray},
  S., {Forman}, W., {Jones}, C., \& {Tucker}, W. 2002, \apjl, 567, L27

\bibitem[{{Marshall}(2006)}]{2006MNRAS.372.1289M}
{Marshall}, P. 2006, \mnras, 372, 1289

\bibitem[{{Marshall} {et~al.}(2002){Marshall}, {Hobson}, {Gull}, \&
  {Bridle}}]{2002MNRAS.335.1037M}
{Marshall}, P.~J., {Hobson}, M.~P., {Gull}, S.~F., \& {Bridle}, S.~L. 2002,
  \mnras, 335, 1037

\bibitem[{{Monaghan}(2005)}]{2005RPPh...68.1703M}
{Monaghan}, J.~J. 2005, Reports of Progress in Physics, 68, 1703

\bibitem[{{Moore} {et~al.}(1999){Moore}, {Ghigna}, {Governato}, {Lake},
  {Quinn}, {Stadel}, \& {Tozzi}}]{1999ApJ...524L..19M}
{Moore}, B., {Ghigna}, S., {Governato}, F., {Lake}, G., {Quinn}, T., {Stadel},
  J., \& {Tozzi}, P. 1999, \apjl, 524, L19

\bibitem[{{Nagamine} {et~al.}(2004){Nagamine}, {Springel}, \&
  {Hernquist}}]{2004MNRAS.348..435N}
{Nagamine}, K., {Springel}, V., \& {Hernquist}, L. 2004, \mnras, 348, 435

\bibitem[{{Natarajan} {et~al.}(2007{\natexlab{a}}){Natarajan}, {De Lucia}, \&
  {Springel}}]{2007MNRAS.376..180N}
{Natarajan}, P., {De Lucia}, G., \& {Springel}, V. 2007{\natexlab{a}}, \mnras,
  376, 180

\bibitem[{{Natarajan} {et~al.}(2007{\natexlab{b}}){Natarajan}, {Kneib},
  {Smail}, {Treu}, {Ellis}, {Moran}, {Limousin}, \&
  {Czoske}}]{2007arXiv0711.4587N}
{Natarajan}, P., {Kneib}, J.-P., {Smail}, I., {Treu}, T., {Ellis}, R., {Moran},
  S., {Limousin}, M., \& {Czoske}, O. 2007{\natexlab{b}}, ArXiv e-prints, 711

\bibitem[{{Natarajan} \& {Refregier}(2000)}]{2000ApJ...538L.113N}
{Natarajan}, P., \& {Refregier}, A. 2000, \apjl, 538, L113

\bibitem[{{Natarajan} \& {Springel}(2004)}]{2004ApJ...617L..13N}
{Natarajan}, P., \& {Springel}, V. 2004, \apjl, 617, L13

\bibitem[{{Okura} {et~al.}(2007){Okura}, {Umetsu}, \&
  {Futamase}}]{2007arXiv0710.2262O}
{Okura}, Y., {Umetsu}, K., \& {Futamase}, T. 2007, ArXiv e-prints, 710

\bibitem[{{Petters} {et~al.}(2001){Petters}, {Levine}, \&
  {Wambsganss}}]{2001stgl.book.....P}
{Petters}, A.~O., {Levine}, H., \& {Wambsganss}, J. 2001, {Singularity theory
  and gravitational lensing} (Singularity theory and gravitational lensing /
  Arlie O.~Petters, Harold Levine, Joachim Wambsganss.~Boston : Birkh{\"a}user,
  c2001.~(Progress in mathematical physics ; v.~21))

\bibitem[{{Rines} {et~al.}(2007){Rines}, {Diaferio}, \&
  {Natarajan}}]{2007ApJ...657..183R}
{Rines}, K., {Diaferio}, A., \& {Natarajan}, P. 2007, \apj, 657, 183

\bibitem[{{Saha} {et~al.}(2001){Saha}, {Williams}, \&
  {Abdelsalam}}]{2001ASPC..237..279S}
{Saha}, P., {Williams}, L.~L.~R., \& {Abdelsalam}, H.~M. 2001, in Astronomical
  Society of the Pacific Conference Series, Vol. 237, Gravitational Lensing:
  Recent Progress and Future Go, ed. T.~G. {Brainerd} \& C.~S. {Kochanek},
  279--+

\bibitem[{{Saha} {et~al.}(2007){Saha}, {Williams}, \&
  {Ferreras}}]{2007ApJ...663...29S}
{Saha}, P., {Williams}, L.~L.~R., \& {Ferreras}, I. 2007, \apj, 663, 29

\bibitem[{{Sand} {et~al.}(2003){Sand}, {Treu}, {Smith}, \&
  {Ellis}}]{2003astro.ph..9465S}
{Sand}, D.~J., {Treu}, T., {Smith}, G.~P., \& {Ellis}, R.~S. 2003, ArXiv
  Astrophysics e-prints

\bibitem[{{Schneider}(1995)}]{1995A&A...302..639S}
{Schneider}, P. 1995, \aap, 302, 639

\bibitem[{{Schneider} {et~al.}(1992){Schneider}, {Ehlers}, \&
  {Falco}}]{1992grle.book.....S}
{Schneider}, P., {Ehlers}, J., \& {Falco}, E.~E. 1992, {Gravitational Lenses}
  (Gravitational Lenses, XIV, 560 pp.~112 figs..~Springer-Verlag Berlin
  Heidelberg New York.~ Also Astronomy and Astrophysics Library)

\bibitem[{{Schneider} \& {Er}(2007)}]{2007arXiv0709.1003S}
{Schneider}, P., \& {Er}, X. 2007, ArXiv e-prints, 709

\bibitem[{{Schneider} \& {Seitz}(1995)}]{1995A&A...294..411S}
{Schneider}, P., \& {Seitz}, C. 1995, \aap, 294, 411

\bibitem[{{Schneider} \& {Weiss}(1992)}]{1992A&A...260....1S}
{Schneider}, P., \& {Weiss}, A. 1992, \aap, 260, 1

\bibitem[{{Seitz} \& {Schneider}(1998)}]{1998astro.ph..2051S}
{Seitz}, S., \& {Schneider}, P. 1998, ArXiv Astrophysics e-prints

\bibitem[{{Seitz} {et~al.}(1998){Seitz}, {Schneider}, \&
  {Bartelmann}}]{1998A&A...337..325S}
{Seitz}, S., {Schneider}, P., \& {Bartelmann}, M. 1998, \aap, 337, 325

\bibitem[{{Springel} \& {Hernquist}(2003)}]{2003MNRAS.339..312S}
{Springel}, V., \& {Hernquist}, L. 2003, \mnras, 339, 312

\bibitem[{{Taylor} {et~al.}(2004){Taylor}, {Bacon}, {Gray}, {Wolf},
  {Meisenheimer}, {Dye}, {Borch}, {Kleinheinrich}, {Kovacs}, \&
  {Wisotzki}}]{2004MNRAS.353.1176T}
{Taylor}, A.~N., {Bacon}, D.~J., {Gray}, M.~E., {Wolf}, C., {Meisenheimer}, K.,
  {Dye}, S., {Borch}, A., {Kleinheinrich}, M., {Kovacs}, Z., \& {Wisotzki}, L.
  2004, \mnras, 353, 1176

\bibitem[{{Voigt} \& {Fabian}(2006)}]{2006MNRAS.368..518V}
{Voigt}, L.~M., \& {Fabian}, A.~C. 2006, \mnras, 368, 518

\bibitem[{{Wilson} {et~al.}(1996){Wilson}, {Kaiser}, {Cole}, \&
  {Frenk}}]{1996AAS...189.8206W}
{Wilson}, G., {Kaiser}, N., {Cole}, S., \& {Frenk}, C. 1996, in Bulletin of the
  American Astronomical Society, Vol.~28, Bulletin of the American Astronomical
  Society, 1385--+

\bibitem[{{Wittman} {et~al.}(2001){Wittman}, {Tyson}, {Margoniner}, {Cohen}, \&
  {Dell'Antonio}}]{2001ApJ...557L..89W}
{Wittman}, D., {Tyson}, J.~A., {Margoniner}, V.~E., {Cohen}, J.~G., \&
  {Dell'Antonio}, I.~P. 2001, \apjl, 557, L89

\bibitem[{{Woolfson}(2007)}]{2007MNRAS.376.1173W}
{Woolfson}, M.~M. 2007, \mnras, 376, 1173

\end{thebibliography}
%\bibliographystyle{apj}

\end{document}